\documentclass[%
 reprint,
 twocolumn
 amsmath,amssymb,
 aps,
]{revtex4-2}

\usepackage{graphicx}% Include figure files
\usepackage{dcolumn}% Align table columns on decimal point
\usepackage{bm}
\usepackage{caption}% bold math
\usepackage{subcaption}
\usepackage{float}
\usepackage{mathtools}% bold math
\usepackage{multirow}
\usepackage{feynmp-auto}
\usepackage{booktabs}
\usepackage{siunitx}
\usepackage{ragged2e}
\DeclarePairedDelimiterX\braket[2]{\langle}{\rangle}{#1 \delimsize\vert #2}

\usepackage{hyperref}  
\hypersetup{
    colorlinks=true,
    linkcolor=blue,
    filecolor=blue,      
    urlcolor=blue,
    citecolor=blue,
    }

\begin{document}

\preprint{APS/123-QED}

\title{Correlated phases and topological phase transition in twisted bilayer graphene at one quantum of magnetic flux}% Force line breaks with \\

\author{ Miguel Sánchez Sánchez}
\email{miguel.sanchez@csic.es}
\author{Tobias Stauber}%
\email{tobias.stauber@csic.es}
\affiliation{Instituto de Ciencia de Materiales de Madrid ICMM-CSIC, Madrid (Spain)}%

% \collaboration{MUSO Collaboration}%\noaffiliation

%  \homepage{http://www.Second.institution.edu/~Charlie.Author}
% \affiliation{
%  Second institution and/or address\\
%  This line break forced% with \\
% }%
% \affiliation{
%  Third institution, the second for Charlie Author
% \affiliation{%
%  Authors' institution and/or address\\
%  This line break forced with \textbackslash\textbackslash
% }%

% \collaboration{CLEO Collaboration}%\noaffiliation


\begin{abstract}

When the perpendicular magnetic flux per unit cell in a crystal is equal to the quantum of magnetic flux, $\Phi_0=h/e$, we enter the 'Hofstadter regime'. The large unit cell of moiré materials like magic-angle twisted bilayer graphene (MATBG) allows the experimental study of this regime at feasible values of the field around $20$ to $30$ T. In this work, we report numerical analysis of a tight-binding model for MATBG at one quantum of external magnetic flux, including the long-range Coulomb and on-site Hubbard interaction. We study the  correlated states for dopings of $-2,0$ and $2$ electrons per unit cell at the mean-field level. We find competing insulators with Chern numbers $2$ and $0$ at positive doping, the stability of which is determined by the dielectric screening, which opens up the possibility of observing a topological phase transition in this system.

\end{abstract}

\maketitle

%\tableofcontents

\section{Introduction}

Magic angle twisted bilayer graphene (MATBG) is a two dimensional quantum material that exhibits a plethora of exotic phases ranging from superconductors to Fractional Chern insulators\cite{Cao2018sup,Yankowitz19,Lu2019,liu21, Jaoui2022,cao20,Wu2021,Stepanov21,Xie2021}. It constitutes a remarkable platform for the understanding of the many-body problem in Condensed Matter and the interplay of strong interactions and topology, and has led to the field of moiré materials\cite{Wang2020,Park2021,scheer2023twistronics,crépel2023chiral}.

Moreover, crystalline systems under magnetic fields are controlled by the scale given by the magnetic flux quantum $\Phi_0 = h/e$\cite{Hofstadter76}. When the field is such that the magnetic flux per unit cell is comparable to $\Phi_0$ (or, equivalently, the magnetic length is comparable to the lattice constant\cite{RevGoerbig}) the different Landau levels merge into Hofstadter bands\cite{wang20,Biao20,Guan22,rodrigues2023atomistic}. In typical materials such magnetic fields are of the order of $10^4$ T, but in MATBG the large moiré unit cell allows to probe the 'Hofstadter regime' by accessible fields of the order of $25$ T. 

In MATBG the Landau level spectrum of the competing correlated states and  has been studied for low magnetic fields\cite{singh2023topological,wang22,Yankowitz19,Lu2019,Wu2021,Stepanov21,Das2021}. Also, at one magnetic flux quantum reentrant correlated insulators have been predicted\cite{herzog22_3} and observed\cite{efetov22}.

% On the theory side, the Bistritzer-McDonald (BM) or continuum model\cite{McDonald11,Peres12} is a low energy theory that has proven very powerful in  understanding the physics of TBG, revealing the emergent symmetries of the system that have led to the picture of the '$U(4)$ ferromagnets' for the correlated insulators\cite{seo19, Po18,Kang19, vafek20, ledwith21,bernevig321,bernevig421}. However, the model, with only a handful of parameters, cannot capture the finer details of the spectrum and the wave functions. These differences at low energy scales are relevant in the competition between states.

% In this work we employ a tight-binding model for MATBG. The high computational cost, which makes atomistic studies scarce in this system\cite{gonzalez20,stauber21,klebl21,Goodwin_2020}, is partially bypassed by a projection onto the subspace of the low energy bands (the 'flat bands'). The external magnetic flux is tuned to zero and one magnetic flux quantum per unit cell, we focus on samples without strain and leave electron-phonon coupling for future work.

% At zero field, the particle-hole asymmetry splits the energy of the valley polarized and Kramers intervalley coherent state, supporting the KIVC at even filling. Previous studies in the BM model established that the KIVC is also supported, but the splitting mechanism is kinetic energy superexchange\cite{Kang19,vafek20,bernevig421,Bultinck20,Kwan23}. The Hubbard interaction makes the spin polarized state competitive at charge neutrality, and 

When the filling is equal to an integer number of electrons per unit cell, correlation induced gaps can arise facilitated by the large interactions compared to the bandwidth of the flat bands. The Hartree-Fock (HF) method has proven effective in capturing the correlated states in MATBG at zero external field\cite{Bultinck20,lin23,adhikari2023strongly,xie20,Kwan23,song21}, mostly in the setting of the Bistritzer-MacDonald continuum model\cite{McDonald11}. We perform self-consistent HF simulations now at one quantum of flux in a microscopic model.

Consistently for different values of the dielectric constant, we observe a Chern insulator with Chern number $-2$ when the doping is of $-2$ electrons per unit cell (denoted by $\nu=-2$). At charge neutrality, a spin-polarized state and a spin-unpolarized insulator are competitive and their stability depends on $\epsilon_r$ and the Hubbard energy $U$. For $\nu=+2$, we observe a topological phase transition from an insulator with Chern number $2$
to an intervalley coherent trivial insulator as we increase the dielectric screening. Experimentally, the data of Ref. \cite{efetov22} shows a correlated insulator and a nearby (competitive) Chern 2 trace for $\nu=+2$. 

 \begin{figure*}[t]
     \centering
     \includegraphics[width=.9\linewidth]{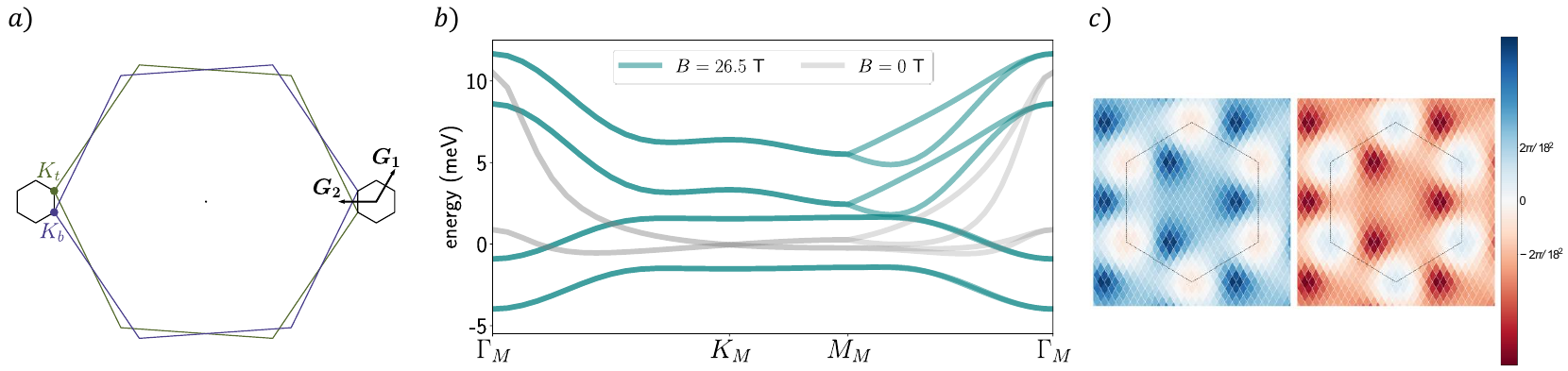}
     \caption{\textbf{a)} The Brillouin zone of the rotated top and bottom layers. $\boldsymbol{K_{b(t)}} = R_{(-)\theta/2}(-4\pi/3a,0)$ are the $K$ points of each layer. The interlayer tunneling couples states with momenta related by the $\boldsymbol{G}$ vectors, producing the Brillouin zone of MATBG. The $K$ and $K'$ regions are very far apart and do not couple to each other. \textbf{b)} Flat bands of MATBG at one flux quantum, in turquoise. The spin-up and down bands are splitted by $\sim 3$ meV due to the Zeeman term. The band structure at $B=0$ T is shown in gray for comparison. \textbf{c)} Integrated Berry curvature on a $18 \times 18$ grid of the $K$ valence band (left) and $K$ conduction band (right). The Chern number is $|C|=1$. The curvature of the $K'$ bands are obtained from $C_{2z}$. The Berry curvatures are almost identical with opposite sign, hinting to an emergent symmetry.}
     \label{bandsnonint}
 \end{figure*}

\section{The model}
Consider two graphene layers stacked on top of each other such that top and bottom atoms are vertically aligned. The bottom layer is rotated by an angle $-\theta/2$, and the top layer by $\theta/2$, with the center of rotation being the center of one of the graphene hexagons. The magic angle sits between $1$ and $1.1^\circ$\cite{Balents2020}. We choose a twist of $\theta = 1.05012^\circ$ that makes the twisted superstructure exactly conmensurate, with lattice constant $L_M = 13.4$ nm and $11908$ atoms in the unit cell.

We employ the Slater-Koster parametrization of the hopping integral $t(\boldsymbol{r})$ of Ref.\cite{Koshino12} with a $p_z$ orbital per carbon atom and spin, giving the tight-binding Hamiltonian 
\begin{align}
    H_0 = \sum_{\boldsymbol{r_i},\boldsymbol{r_j},s} t(\boldsymbol{r_i}-\boldsymbol{r_j}) c^{\dagger}_{\boldsymbol{i}s}c_{\boldsymbol{i}s}, %\nonumber \\
    \end{align}
$c^\dagger_{\boldsymbol{i},s}$ being the creation operator of an electron with spin $s$ at position $\boldsymbol{r_i}$.  Details on the geometry of MATBG and the hopping parameters can be found in Appendix \ref{appendixa}. 
% The hopping integral is decomposed into $\sigma$ and $\pi$-bond hoppings,
% \begin{align}
% % \end{equation}
% % \begin{equation}
%     t(\boldsymbol{r}) = - &V_{pp\pi}(r) \Bigg(1 - \bigg(\frac{\boldsymbol{r}\cdot \boldsymbol{\hat{z}}}{ r}\bigg)^2\Bigg) + V_{pp\sigma}(r) \bigg(\frac{\boldsymbol{r}\cdot \boldsymbol{\hat{z}}}{ r}\bigg)^2, \nonumber \\
%     &V_{pp\pi}(r) = V_{pp \pi}^0 e^{-(r - a_0)/r_0}, \nonumber \\ 
%     &V_{pp\sigma}(r) = V_{pp \sigma}^0 e^{-(r - d_0)/r_0}, 
% \end{align}
% with the parameters $V^0_{ppp\pi} = 2.7$ eV,  $V^0_{pp\sigma} = 0.48$ eV and $r_0 = 0.0453 $ nm. 
The Zeeman energy reads
\begin{align}
    H_Z = - \frac{g\mu_B B }{2} \sum_{\boldsymbol{r_i}}c_{\boldsymbol{i}\uparrow}^\dagger c_{\boldsymbol{i} \uparrow} - c_{\boldsymbol{i}\downarrow}^\dagger c_{\boldsymbol{i} \downarrow},
\end{align}
with $g=2$ the gyromagnetic ratio of the electron and $\mu_B$ the Bohr magneton.

The electrons interact through the double-gated Coulomb potential
\begin{align}
    V = \frac{1}{2}\sum_{\boldsymbol{r_i} \neq \boldsymbol{r_j} s_i s_j} V(\boldsymbol{r_i}-\boldsymbol{r_j}) :c^\dagger_{\boldsymbol{i},s_i} c_{\boldsymbol{i}, s_i} c^\dagger_{\boldsymbol{j},s_j} c_{\boldsymbol{j}, s_j}:, \nonumber \\
    V(\boldsymbol{r_i}-\boldsymbol{r_j}) = \frac{e^2}{4\pi \epsilon_0 \epsilon_r}\sum_{n} \frac{(-1)^n}{||\boldsymbol{r_i} - \boldsymbol{r_j} + n\xi \boldsymbol{\Hat{z}}||},
    \label{potential}
 \end{align}
which applies for the experimental setups where two metallic plates are placed at $z=\pm \xi/2$. We set $\xi= 10$ nm throughout the paper. The dielectric constant $\epsilon_r$ accounts for the screening due to the substrate and internal screening due to the electrons. The interaction is normal ordered\cite{giuliani_vignale_2005} with respect to the ground state of two decoupled graphene layers at charge neutrality. This choice of normal ordering is also called graphene subtraction scheme\cite{xie20,lin23}. In the calculation of the decoupled ground state we have not included the Zeeman splitting.
% Under magnetic field, we do not include the Zeeman shift when calculating the graphene state, so that the spin imbalances come entirely from the flat band physics. 
The on-site Hubbard term is also considered,
\begin{align}
 H_U = U \sum_{\boldsymbol{r_i}} :c^\dagger_{\boldsymbol{i}\uparrow} c_{\boldsymbol{i}\uparrow} c^\dagger_{\boldsymbol{i}\downarrow} c_{\boldsymbol{i}\downarrow}:.
 \label{hubbard}
\end{align}
It can be thought of as a regularization of the Coulomb potential at $\boldsymbol{r} = \boldsymbol{0}$.

The total Hamiltonian is then $H = H_0 + H_Z + V + H_U$.
 
At zero flux, the point group of MATBG is $D_6$, generated by six-fold rotations around the $z$ axis, $C_{6z}$, and two-fold rotations around the $y$ axis, $C_{2y}$, leaving the origin fixed (below, we will be adressing the rotations $C_{3z} = C_{6z}^2$ and $C_{2z}=C_{6z}^3$). The spin-orbit coupling being small, spinless time-reversal $\mathcal{T}$ is also a symmetry.
Under magnetic flux, the time reversal $\mathcal{T}$ and rotations $C_{2y}$ reverse the sign of the external field, and only the combined $C_{2y}\mathcal{T}$ is preserved. On the other hand, the rotations around the $z$ axis are preserved\cite{Herzog22}.

\subsection*{Minimal coupling to the external magnetic field}

At nonzero magnetic field, the Peierls' substitution\cite{Luttinger51} adds a phase to the hopping elements,
\begin{align}
    t(\boldsymbol{r_i}-\boldsymbol{r_j}) \to  t(\boldsymbol{r_i}-\boldsymbol{r_j}) e^{i\theta_ {\boldsymbol{i},\boldsymbol{j}}}, \nonumber \\
    \theta_{\boldsymbol{i},\boldsymbol{j}} = \frac{2\pi}{\Phi_0} \int_{\boldsymbol{r_i}\to \boldsymbol{r_j}} \boldsymbol{A}(\boldsymbol{r'}) \cdot d\boldsymbol{r'},
\end{align}
where $\Phi_0 = h/e$ is the quantum of magnetic flux, and the line integral goes from $\boldsymbol{r_i}$ to $\boldsymbol{r_j}$ in a straight line if the basis orbitals are well localized\cite{Biao20}. \\
In the presence of magnetic flux, the translation operators pick up a phase. They act on the single-particle states as\cite{Herzog20}
\begin{align}
    \tilde{T}_1 &= \sum_{\boldsymbol{r_i}} e^{-2\pi i \xi_{2\boldsymbol{i}}\phi - i\theta_{\boldsymbol{i},\boldsymbol{i}+\boldsymbol{L_1}}} c^\dagger_{\boldsymbol{i}+\boldsymbol{L}_1}c_{\boldsymbol{i}}, \nonumber \\
    \tilde{T}_2 &= \sum_{\boldsymbol{r_i}} e^{2\pi i \xi_{1\boldsymbol{i}}\phi - i\theta_{\boldsymbol{i},\boldsymbol{i}+\boldsymbol{L_2}}} c^\dagger_{\boldsymbol{i}+\boldsymbol{L_2}}c_{\boldsymbol{i}},
\end{align}
where $\xi_{\boldsymbol{i}1,2}$ are defined from the lattice vectors $\boldsymbol{L}_{1,2}$ by $\boldsymbol{r_i} = \xi_{\boldsymbol{i}1} \boldsymbol{L}_1 + \xi_{\boldsymbol{i}2} \boldsymbol{L}_2$ and  $\phi = \Phi / \Phi_0 = BA_M/\Phi_0$ is the flux per moiré unit cell in units of $\Phi_0$.

It can be shown that $[\mathcal{H}, \tilde{T_1}] = [\mathcal{H}, \tilde{T_2}] = 0$ and $\tilde{T}_1\tilde{T}_2 = e^{-2\pi i \phi }\tilde{T}_2 \tilde{T}_1$\cite{Herzog22}, so the translational symmetries are broken in general. However, if $\phi$ is a rational number $p/q$ one can choose the set of commuting operators $( \tilde{T}_1, \tilde{T}_2^q)$, or $(\tilde{T}_1^q, \tilde{T}_2)$, and diagonalize them simultaneously with the Hamiltonian. Translational symmetry is then recovered at rational fluxes with a unit cell that is $q$ times larger than at zero flux, and the Bloch waves are generalized to magnetic waves having good $\tilde{T}_1$ and $\tilde{T}_2^q$ quantum numbers. 

In the periodic Landau gauge\cite{Cuniberti07} the vector potential reads
\begin{align}
    &\boldsymbol{A}(\boldsymbol{r}) = \frac{\Phi}{2\pi}\Bigg(\xi_1 \boldsymbol{G}_2 - 2\pi \boldsymbol{\nabla}\big(\xi_2\left \lfloor{\xi_1 + \epsilon}\right \rfloor\big)  \Bigg) \nonumber \\
    &= \frac{\Phi}{2\pi} \Bigg(-\xi_2\sum_{n=-\infty}^{\infty}\delta(\xi_1 - n + \epsilon)\boldsymbol{G}_1 + (\xi_1 - \left \lfloor{\xi_1 + \epsilon}\right \rfloor) \boldsymbol{G}_2\Bigg), 
\end{align}
with $\boldsymbol{G}_{1,2}$ the reciprocal vectors and $\lfloor ... \rfloor$ the floor function. In this gauge the phases of the translation operators $\tilde{T}_2^q,$ $\tilde{T}_1$ cancel and the Bloch waves have the same form as in zero flux. The infinitesimal $\epsilon$ prevents ambiguities in the Peierls' phases if some atoms lie at integer values of $\xi_1$. The momentum $\boldsymbol{k}$ takes the possible values in the magnetic Brillouin zone of the dual lattice with lattice vectors $\boldsymbol{G}_1$ and $\boldsymbol{G}_2/q$.

In our case of interest, for MATBG the unit flux magnetic field depends on the twist angle as $B\approx 24.048\ \theta(^\circ)^2$ T, giving $B=26.51$ T and a Zeeman splitting of $\mu_B B = 1.535$ meV for $\theta=1.05^\circ$.

\begin{figure}[b]
    \centering
    \includegraphics[width=.8\linewidth]{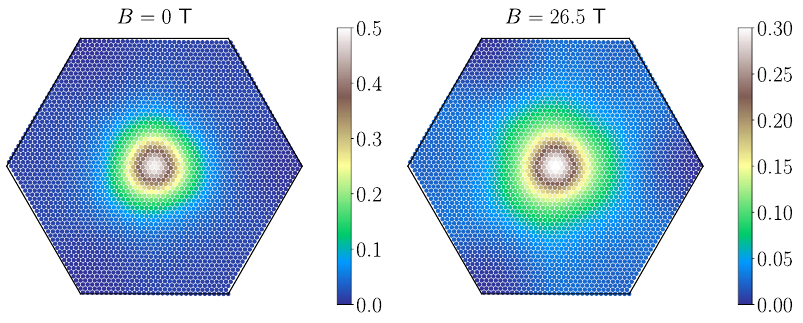}
    \caption{\textbf{Density distribution of the flat bands}, for $0$ and $26.5$ T, in arbitrary units. We plot the density in the bottom layer and sublattice $A$, while the density in the remaining sublattices and layers can be obtained by symmetry. It is centered in the $AA$ region of the unit cell in both cases, with different spreads.}
    \label{dens}
\end{figure}

\subsection*{Non interacting band structure}

In Fig. \ref{bandsnonint}b) we plot the band structure of MATBG at $26.51$ T along the $\Gamma_M K_M M_M \Gamma_M$ line. The crystal momentum is not gauge invariant, and at nonzero flux the position of the high symmetry points is shifted with respect to their locations at zero flux. We discuss this further in Appendix \ref{appb}.

The almost exact degeneracies along $\Gamma_M K_M M_M$ are due to the negligible scattering between the two valleys of the monolayers of graphene, so that the valley is a good quantum number, see Appendix \ref{appf}. The valley charge commutes with $C_{3z}$ and $C_{2y}\mathcal{T}$ and anticommutes with $C_{2z}$.

\begin{figure}[b]
    \centering
    \includegraphics[width=.9\linewidth]{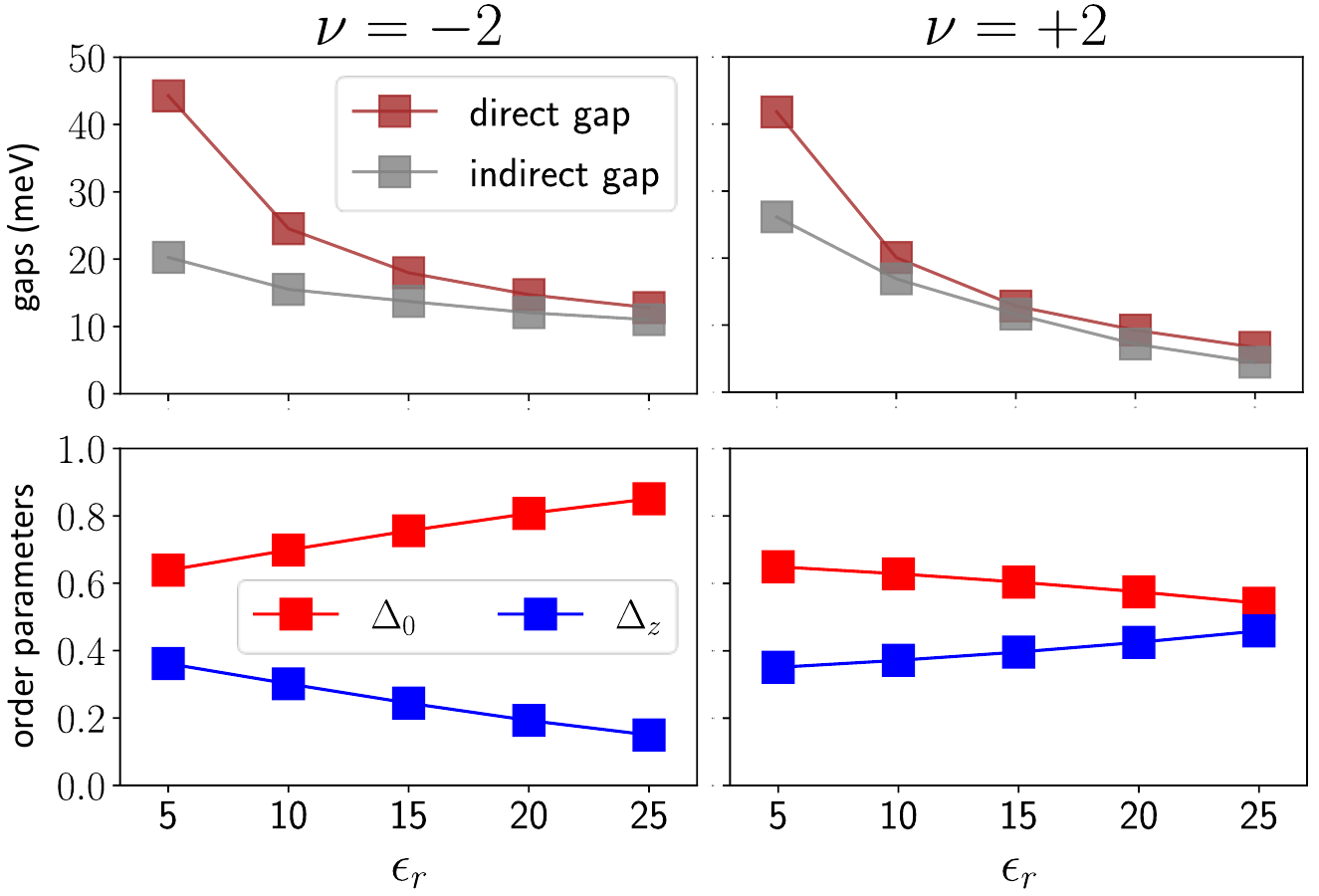}
    \caption{\textbf{Gaps and order parameters for $\boldsymbol{\nu=\pm 2}$}. The direct gap is obtained for transitions between states with the same spin, and the indirect gap is the gap in the total density of states. The order parameters are defined in Eq. \ref{oparams}. $U$ was set to $4$ eV.}
    \label{gapsandords}
\end{figure}

Also, the Dirac cones are gapped due to the breaking of $C_{2z}\mathcal{T}$\cite{junyeong19}. The gap at the $K$ points is about $5$ meV. This is in contrast to MATBG at zero flux, where the bands are very flat with a bandwidth of about $1$ meV, except only at the $\Gamma_M$ point\cite{vafek23}. The Zeeman splitting of $3.07$ meV is comparable to the bandwidth.

Regarding the topology, we have computed the action of the rotations on the Bloch states at the high-symmetry momenta. The $C_{3z}$ eigenvalues are $((\omega,\omega),(\omega,\omega))$ at $\Gamma_M$,  $((\omega^*,\omega^*),(1,1))$ at $K_M$ and $((\omega^*,\omega^*),(1,1))$ at $K'_M$, where $\omega = e^{2\pi i/3}$ and the first parenthesis refers to the valence bands and the second to the conduction bands. It follows that the valence bands have a Chern number of $-1$ mod $3$ and the conduction bands of $+1$ mod $3$\cite{fang12}. $C_{2z}$ acts as the Pauli $x$ matrix on the doublets at $\Gamma_M$ and $M_M$. This is due to the fact that there is one state from each graphene valley in the doublets. 

Following the theory of topological quantum chemistry\cite{Bradlyn2017}, we infer that the flat bands are topologically trivial and can be Wannierized keeping the valley symmetry manifest. For each valley sector, the flat bands can be constructed from two Wannier orbitals with $C_{3z}$ eigenvalue $\omega$ centered at the Moiré zone corners (the AB and BA sites) and related by $C_{2y}\mathcal{T}$. As shown in Fig. \ref{dens}, the density profile of the flat bands is centered around the AA-stacked region. This forces the Wannier orbitals to exhibit a three-peak structure similarly to MATBG at zero flux \cite{kang2018,zang2022real}.

 \begin{figure*}[t]
     \centering
     \includegraphics[width=.99\linewidth]{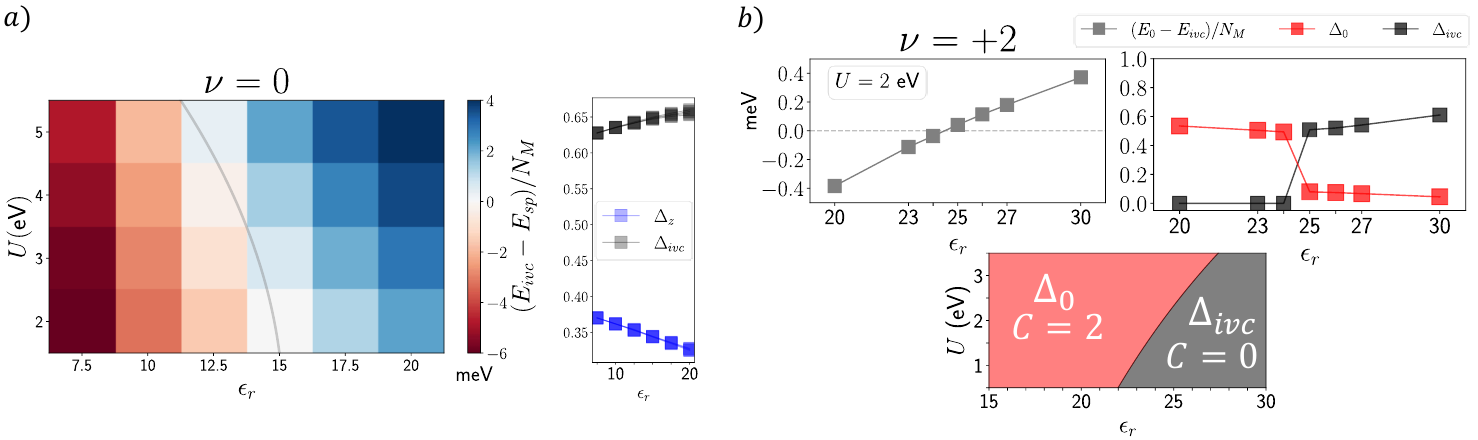}
     \caption{\textbf{a) Competing states at charge neutrality.} Left: The difference in the energy of the intervalley coherent ($E_{ivc}$) and spin polarized ($E_{sp}$) states per unit cell as a function of $\epsilon_r$ and $U$. A tentative phase boundary is drawn in gray. Right: The order parameters $\Delta_{ivc}$ and $\Delta_{z}$ (see the text for their definitions) of the intervalley coherent (ivc) solution as a function of $\epsilon_r$ for $U=2,3,4$ and $5$ eV. The curves for different $U$ are almost identical and fall on top of each other. \textbf{b) The phase transition for $\boldsymbol{\nu=+2}$}. Top: We plot the difference in energy per unit cell of the self-consistent state with dominant order $\Delta_0$ and the ivc state to the left, and the order parameters of the ground state to the right. $\Delta_z$ is of order $0.3-0.4$ in both sides of the transition. $U$ was set to $2$ eV. Bottom: the $\epsilon_r-U$ phase diagram with the dominant order parameters and Chern numbers, showing the phase transition.} 
     \label{resultsnu0+2}
 \end{figure*}
 
\subsection*{The irrep basis}

The 'irrep' basis of the flat bands is defined by the action of the 'particle-hole' operator, $C_{2z}P$\cite{herzog22_2, herzog22_3, Bultinck20, bernevig321}. $C_{2z}P$ is local in momentum space, unitary, hermitian and squares to $1$. Like the valley charge, it is an emergent operator at low energies. The implementation of both operators on the lattice is detailed in Appendix \ref{appc}. Let us remark that in our tight-binding model this operator has a generic form in the $H_0$ eigenbasis. Contrarily to Refs. \cite{herzog22_2,herzog22_3,Bultinck20,bernevig321}, it is not strictly off-diagonal in the band basis (hence the name particle-hole coined there).

In certain limit, $C_{2z}P$ is the generator of a symmetry of the model that adds to the usual valley charge conservation of MATBG, see Appendix \ref{appd} for a discussion of the symmetries and their breaking.

In the irrep gauge we have
\begin{align}
    [\overline{C_{2z}P}(\boldsymbol{k})]_{\eta\lambda,\eta'\lambda'} = \langle \boldsymbol{k}\eta \lambda | C_{2z}P | \boldsymbol{k}\eta'\lambda' \rangle = [\tau_x]_{\eta \eta'} [\lambda_0]_{\lambda  \lambda'},
    % \overline{C_{2z}P}(\boldsymbol{k})\Big(\overline{C_{2z}P}(\boldsymbol{k})\overline{C_{2z}P}(\boldsymbol{k})^\dagger \Big)^{-1/2} = \tau_x \lambda_0.
    \label{c2pp1in}
\end{align}
$|\boldsymbol{k} \eta \lambda \rangle$ denoting a Bloch state with momentum $\boldsymbol{k}$, valley $\eta$ and irrep number $\lambda$ ($\eta, \lambda =\pm 1$). Valley $K$ will be associated with $\eta=+1$, and valley $K'$, or $-K$, with $\eta=-1$. $\tau_{0,xyz}$ and $\lambda_{0,xyz}$ are the identity and Pauli matrices in valley and irrep number space, respectively. Here we omit the spin index, keeping in mind that we construct one copy of the irrep basis for each spin.

Actually, the the singular values of the projected matrix $\overline{C_{2z}P}(\boldsymbol{k})$ (plotted in Appendix \ref{appc}) are close but not equal to $1$. Hence we must modify Eq. \ref{c2pp1in} to 
\begin{align}
    % \langle \boldsymbol{k}\eta \lambda | C_{2z}P | \boldsymbol{k}\eta'\lambda' \rangle = [\tau_x]_{\eta \eta'} [\lambda_0]_{\lambda  \lambda'}.
    \overline{C_{2z}P}(\boldsymbol{k})\Big(\overline{C_{2z}P}(\boldsymbol{k})\overline{C_{2z}P}(\boldsymbol{k})^\dagger \Big)^{-1/2} = \tau_x \lambda_0,
    \label{c2pp1}
\end{align}
where the inverse square root makes the matrix unitary.
% In the continuum model, the chiral limit is topologically distinct from the real system at one flux quantum\cite{herzog22_2}, a phenomenon that is reflected in the tight-binding model. In this case, the irrep basis is not maximally polarized, hence the sublattice polarization and irrep character cannot be simultaneously manifested.

We further fix the  $C_{2z}$ phase to
\begin{align}
\langle [\boldsymbol{-k}] \rho | C_{2z} | \boldsymbol{k} \rho'\rangle= -[\tau_y \lambda_y]_{\rho \rho'},
\end{align}
with $\rho$ the multi-index for valley and irrep, and $[\boldsymbol{k}]$ the momentum $\boldsymbol{k}$ translated to inside the Brillouin zone. 

Finally, notice that the irrep basis is only defined up to arbitrary transformations $V(\boldsymbol{k})$ in both valleys
\begin{align}
    &| \boldsymbol{k}\eta \lambda \rangle \longrightarrow [V(\boldsymbol{k})]_{\lambda \lambda'} | \boldsymbol{k} \eta \lambda' \rangle, \nonumber \\
    &V^\dagger([-\boldsymbol{k}])\lambda_y V(\boldsymbol{k}) = \lambda_y.
    \label{irrepv}
\end{align}

\section{Hartree-Fock results}

We have carried out self-consistent HF simulations of MATBG projected onto the subspace of the flat bands. We describe the HF formalism and the flat band projection method in Appendix \ref{appe}. We remind the reader at this point that there are $8$ flat bands in total ($2$ per valley per spin), and the doping is parametrized by $\nu \in (-4,4)$, where $\nu=0$ denotes the charge neutrality point.   

% Typical values for $\epsilon$ found in the literature range from about $7$ to $12$\cite{Zhang22,Bultinck20}, so we choose $\epsilon=10$ and a realistic value for $U$ of $4$ eV\cite{stauber21,jimenopozo2023short}. However, it has been argued that internal screening is large in these systems and a more appropriate value for $\epsilon$ is several times larger\cite{stauber21,Gonzalez23}. This agrees with the fact that lower values of $\epsilon$ overestimate the gap of the insulators, which in transport are found to be $\lesssim$ 1 meV\cite{Lu2019,Yankowitz19,liu21,efetov22,Pierce2021}. We account for both scenarios and report results also for $\epsilon=50$ and $0.5$ eV.

The self-consistent state $|\text{GS}\rangle$ is characterized by the $Q$ matrix, defined by 
\begin{align}
    [Q(\boldsymbol{k})]_{\rho \rho'} &= 2[P(\boldsymbol{k})]_{\rho \rho'} - \delta_{\rho\rho'}, \nonumber \\
    [P(\boldsymbol{k})]_{\rho \rho'} &= \langle \text{GS} | d^\dagger_{\boldsymbol{k}\rho} d_{\boldsymbol{k}\rho'}| \text{GS} \rangle,
\end{align}
where $d^\dagger_{\boldsymbol{k}\rho}$ creates an electron in state $|\boldsymbol{k}\rho \rangle$. It has the properties $Q(\boldsymbol{k}) = Q(\boldsymbol{k})^\dagger$, $Q(\boldsymbol{k})^2 = 1$ and $\text{tr}(Q(\boldsymbol{k})) = 2\nu$.

In the self-consistent loop we restrain $Q$ to be diagonal in spin so we have $Q(\boldsymbol{k}) = Q_{\uparrow} (\boldsymbol{k}) P_{\uparrow} + Q_\downarrow (\boldsymbol{k}) P_{\downarrow}$ ($P_{s}$ is the projector onto spin $s$). Furthermore, if one of the spin projections is half-filled, $Q_s$ can be expressed as a linear combination of products of Pauli matrices,
\begin{align}
    Q_s(\boldsymbol{k}) 
    =  \mathop{\sum_{\alpha, \beta}} A^s_{\alpha \beta}(\boldsymbol{k})  \lambda_\alpha \tau_\beta,
\end{align}
with real coefficients $A^s_{\alpha,\beta}(\boldsymbol{k})$ and $\sum_{\alpha \beta} (A^s_{\alpha \beta}(\boldsymbol{k}))^2 = 1 $ (and additional constraints to satisfy $Q^2(\boldsymbol{k})=1$).

As stated above, the dielectric constant $\epsilon_r$ in Eq. \ref{potential} depends on the external substrate and the internal screening. Moreover, it will in general depend on $\boldsymbol{r}$, or equivalently on the momentum transfer $\boldsymbol{q}$. In constrained random-phase approximation (RPA) calculations the static dielectric function at zero magnetic field varies between about $10$ and $20$\cite{pizarro19,vanhala20}. 
% An accurate treatment of the dynamical screening
% % , e.g. with the GW approximation\cite{hedin99,marie2023gw},
% might be necessary for the study of MATBG.
Here we take $\epsilon_r$ as a model parameter and perform the self-consistent simulations as a function of $\epsilon_r$. The Hubbard energy $U$ can also vary between $2$ and $5$ eV (the value of $U$ is thought to be $\sim 4$ eV \cite{jimenopozo2023short}).

The self-consistent states do not break the translational (which is imposed) or point symmetries, but they show interesting features in the spin, valley and particle-hole spaces. We report our findings below.

\subsection*{$\boldsymbol{\nu = \pm 2}$}
We find gapped states at electron and hole doping for a wide range of interaction strengths.
% values of $\epsilon_r$. The value of $U$ only modifies weakly the self-consistent state, as can be seen in Fig. \ref{resultsnu0+2}a) for the case $\nu=0$, so we will show here the results for $U=4$ eV.
These insulators are maximally spin polarized in the spin up direction, i.e. at $\nu=-2$ there are two spin up filled bands and at $\nu=+2$ there are four spin up and two spin down bands. The spin polarization stems from the dynamics of the Coulomb interaction, similarly to the zero field case \cite{adhikari2023strongly, Bultinck20, bernevig421}, and the Zeeman term only selects the up direction of the total spin.  

The dominant order parameters are $\sum_{i=x,y,z}A^s_{iz}(\boldsymbol{k})^2$ and $\sum_{i=x,y,z}A^s_{i0}(\boldsymbol{k})^2$, with $s$ the half-filled spin projection. Notice that because of the gauge ambiguity of Eq. \ref{irrepv}, only the above sums of squares result in gauge invariant order parameters. In Fig. \ref{gapsandords} we plot the many body gaps as well as the integrated quantities
\begin{align}
    \Delta_z = \frac{1}{N_M}\sum_{\boldsymbol{k}}\sum_{i=x,y,z} A^s_{iz}(\boldsymbol{k})^2, \nonumber \\
    \Delta_0 = \frac{1}{N_{M}}\sum_{\boldsymbol{k}}\sum_{i=x,y,z} A^s_{i0}(\boldsymbol{k})^2,
    \label{oparams}
\end{align}
for different values of $\epsilon_r$ and $U=4$ eV. $N_M$ is the number of unit cells or, equivalently, the number of $\boldsymbol{k}$ points in the Brillouin zone. 

The solutions exhibit very weak dependence on $U$ for fixed $\epsilon_r$. This is well exemplified in Fig. \ref{resultsnu0+2}a), where the order parameters show negligible dependence on $U$. Although there the results correspond to $\nu=0$, the same phenomenon appears at $\nu=\pm 2$. Finally, the Chern numbers are $C=-2$ for $\nu=-2$ and $C=2$ for $\nu=+2$.

\subsection*{$\boldsymbol{\nu=0}$}

There are two fixed points of the HF numerics for $\nu=0$, one of them being spin polarized (sp) and the other spin-unpolarized with the same wavefunction for the two spin projections. The spin-unpolarized state exhibits intervalley coherent (ivc) order $A^s_{0y}(\boldsymbol{k})$, where the two valleys are in superposition in the many-body wave function. The corresponding integrated order parameter is defined as
\begin{align}
    \Delta_{ivc} = \frac{1}{N_M}\sum_{\boldsymbol{k}} A^s_{0y}(\boldsymbol{k})^2.
\end{align}
Under a transformation of the $U(1)$ valley symmetry of angle $\phi$ acting as $|\boldsymbol{k}\eta \lambda \rangle \to e^{i\eta \phi}|\boldsymbol{k}\eta \lambda \rangle$, the coefficients transform as $A^s_{0y}(\boldsymbol{k}) \to \cos(2\phi)A^s_{0y}(\boldsymbol{k}) + \sin(2\phi) A^s_{0x}(\boldsymbol{k})$ and $A^s_{0x}(\boldsymbol{k}) \to \cos(2\phi)A^s_{0x}(\boldsymbol{k}) - \sin(2\phi) A^s_{0y}(\boldsymbol{k})$, hence the valley symmetry is spontaneously broken in this phase.

In Fig. \ref{resultsnu0+2} we depict the $\epsilon_r-U$ phase diagram and the order parameters of the ivc phase. For most of the phase diagram the ground state is gapped with Chern number $0$, except only when the sp state is metallic (see Appendix \ref{appf}).

\subsection*{Topological phase transition for $\boldsymbol{\nu=+2}$}

We find an intervalley coherent solution for dielectric constants greater than 20. In Fig. \ref{resultsnu0+2}b) we plot the energy difference between the ivc insulator and the Quantum Hall state that is stable for lower screening and the main order parameter around the transition, with $U$ set to $2$ eV. This is a topological transition with a change in Chern number of $2$. The data can be extrapolated with a good accuracy to other values of $U$, showing a critical screening $\epsilon_r^*$ of 
\begin{align}
    \frac{1}{\epsilon_r^*} = \frac{1}{24.4} - 0.003 \big( U(\text{eV})-2 \big).    
\end{align}

\section{Discussion}

In this work we have studied the mean-field phases of MATBG in the Hofstadter regime, at $26.5$ T of external perpendicular magnetic field. We have used an atomistic model for MATBG, which provides precise band structures and wave functions. The flat bands are topologically trivial and can be Wannierized keeping the valley symmetry manifest. The Wannier orbitals extend to neighbouring unit cells, which forces any interacting model of the flat bands to have extended interactions.

We focus on even fillings of $-2,0,2$ electrons per unit cell. The order parameters of the correlated states depend on the values of the dielectric constant $\epsilon_r$ and Hubbard energy $U$. In our case these are model parameters, but the true values may be computed with some method that treats accurately the screening, e.g. the GW approximation\cite{hedin99,marie2023gw}. Another parameter of the model is the reference state chosen as a subtraction point to avoid double counting of the interactions\cite{xie20}. Several subtraction schemes have been used in the literature for $B=0$ T\cite{Kwan23,Bultinck21}. Such choice may influence the results, in particular the breaking (or not) of the $C_{2z}P$-generated symmetry. Our findings reveal the existence of multiple competing states in systems with large symmetry groups like MATBG, and highlight the importance of carrying an exhaustive search for symmetry-breaking patterns in the numerics\cite{Bultinck20,christos22}, specially in the atomistic models where the symmetries are only emergent \cite{sánchez2023correlated}.

In Ref. \cite{efetov22} the authors perform transport measurements on MATBG at one quantum of external magnetic flux. The find a correlated insulator state for $\nu=+2$ and a highly resistive phase that extends form $\nu=-2$ to charge neutrality. The nature of the phase for $\nu=-2,0$ is elusive and cannot be captured by our Hartree-Fock method.

We compare now our results with the experimental data for $\nu=+2$. Besides the correlated insulator, there is a nearby Chern $2$ trace that converges to the point $(\nu=+2,\Phi=\Phi_0)$ and is supressed only very close to that point. We speculate that the intervalley coherent state of our simulations corresponds to the insulator observed in Ref. \cite{efetov22}, while our Quantum Hall state is the supressed $C=2$ insulator in the experimental phase diagram. We comment that the intervalley coherence can be detected as a Kekule pattern on the graphene scale in the scanning tunnelling microscopy (STM) signal\cite{Nuckolls2023-cb}. 

In light of our results, we propose that the manipulation of the screening, either via dielectric engineering\cite{pizarro19,liu21} or by changing the metallic gate distance\cite{Stepanov2020} can induce the topological phase transition from the intervalley coherent insulator to the Chern insulator. We notice that our results are intrinsically in weak coupling, as the metallic plate distance is set to $\xi=10$ nm whereas in Ref. \cite{efetov22} $\xi = 20-30$ nm was used. Alternatively, manipulating the bandwidth and hence modifying the interaction strength relative to kinetic energy, either by hydrostatic pressure\cite{Yankowitz19} or twist angle engineering is another possibility for observing this phase transition.

\section{Acknowledgements}
This work has been supported by MICINN (Spain) under Grants No. PID2020-113164GBI00 and PRE2021-097070. The access to computational resources of Centro de Supercomputación de Galicia (CESGA) is also gratefully acknowledged.

% \clearpage
\bibliography{bibliography}%

\onecolumngrid
\appendix
\setcounter{figure}{0}
\renewcommand\thefigure{\thesection.\arabic{figure}}    

\section{Geometry of MATBG and tight-binding parameters}\label{appendixa}
In graphene, the primitive vectors are $\boldsymbol{a}_1 = a(1/2,\sqrt{3}/2)$ and $\boldsymbol{a}_2 = a(-1/2,\sqrt{3}/2)$, with $a = \sqrt{3}a_0$ and $a_0=0.142$ nm the carbon-carbon distance.  Atoms at lattice points belong to sublattice $A$, and their nearest neighbours displaced by $(\boldsymbol{a}_1 + \boldsymbol{a}_2)/3$  to sublattice $B$. 

Consider two graphene layers stacked on top of each other, at $z=-d_0/2$ and $z=d_0/2$ respectively, being $d_0 = 0.335$ nm the interlayer distance, such that top and bottom atoms are vertically aligned. The bottom layer is rotated by an angle $-\theta/2$, and the top layer by $\theta/2$, with the center of rotation being the center of one of the graphene hexagons. We choose a value of $\theta$ that makes the twisted structure commensurate\cite{Peres12}. In our case, we parametrize the angle by an integer $n$ such that $\cos(\theta) = 1 - 1/2(3n^2+3n+1)$. The unit vectors of the superlattice are \begin{align}
\boldsymbol{L}_1 &= R_{-\theta/2}\big(n\boldsymbol{a}_1 + (n+1)\boldsymbol{a}_2\big) = L_M (0,1), \nonumber \\
\boldsymbol{L}_2 &= R_{\pi/3}\boldsymbol{L}_1
 = R_{-\theta/2}\big((-n-1)\boldsymbol{a}_1 + \big(2n+1)\boldsymbol{a}_2 \big),
\end{align} with $R_{\alpha}$ a rotation by  angle $\alpha$ and $L_M$ the lattice constant. The reciprocal vectors are given by 
\begin{align}
    a_0 \boldsymbol{G}_1 &= G_\theta R_{-\theta/2}\big((3n+1)\boldsymbol{a}_1+\boldsymbol{a}_2\big),  \nonumber \\
   a_0 \boldsymbol{G}_2 &= R_{-2\pi/3}(\boldsymbol{G}_1) \nonumber \\ &= G_\theta R_{-\theta/2}\big(-(3n+2)\boldsymbol{a}_1+(3n+1)\boldsymbol{a}_2\big) ,
\end{align}
where $G_\theta = \frac{4\pi}{3a_0}(9n^2+9n+3)^{-1}$. The magic angle is approximately given by $n=31$ ($1.05^\circ$), corresponding to a Moiré lattice constant of $L_M = 13.4$ nm and $11908$ atoms in the unit cell.

\begin{figure}[H]
    \centering
    \hspace{-4.5cm} \large{$a)$} \hspace{6.5cm}   $b)$ \\
    \includegraphics[width=.3\linewidth]{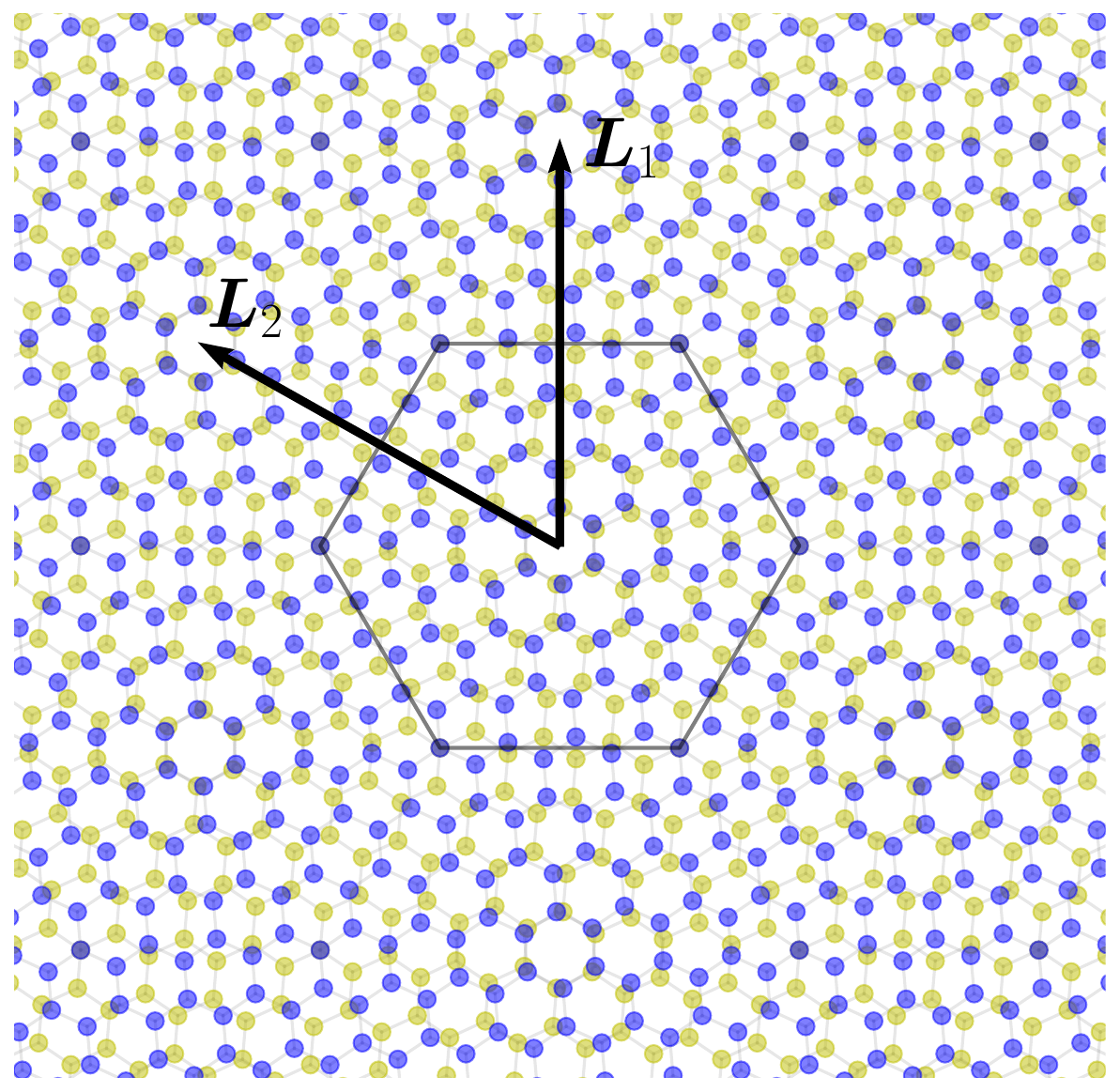} 
    \hspace{1cm}
    \includegraphics[width=.3\linewidth]{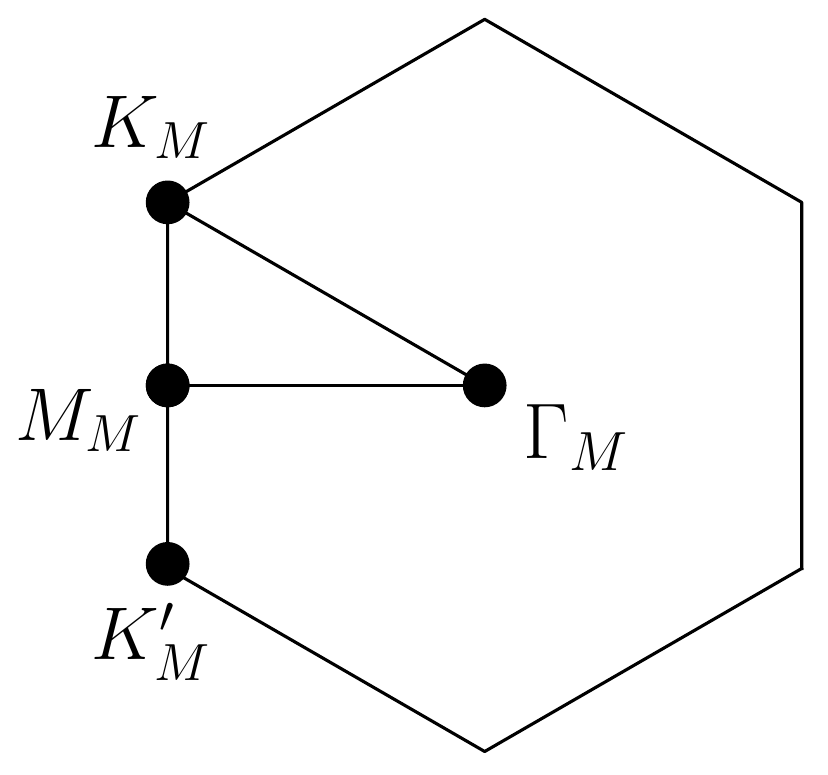}
    \caption{\textbf{a)} Top view of MATBG for a twist angle of $9.43^\circ$. The lattice vectors $\boldsymbol{L}_1$ and $\boldsymbol{L}_{2}$, and the Wigner-Seitz cell are marked. The center of the unit cell is locally $AA$ stacked (vertical alignment of the layers), while at the corners the stacking is locally $AB$ or $BA$ ($A$ atoms of one layer on top of $B$ atoms of the other). \textbf{b)} The Brillouin zone of MATBG with the high symmetry points, $\Gamma_M$, $K_M$, $K'_M$ and $M_M$ labeled. The lines $\Gamma_M K_M$ and $M_M \Gamma_M$ that are considered in the band structure plots are also depicted.}
    \label{lattice}
\end{figure}

Lattice relaxation is included via in-plane distortions following the model of Ref.\cite{Koshino17}. The effect of relaxation is to enlarge the AB and BA regions and reduce the AA regions of the Moiré pattern (see Fig. \ref{lattice}), preserving all the crystallographic symmetries.

We employ the Slater-Koster parametrization of the hopping integral of Ref.\cite{Koshino12}, with a $p_z$ orbital per carbon atom and spin. The hopping integral is decomposed into $\sigma$ and $\pi$-bond hoppings,
\begin{align}
% \end{equation}
% \begin{equation}
    t(\boldsymbol{r}) = - &V_{pp\pi}(r) \Bigg(1 - \bigg(\frac{\boldsymbol{r}\cdot \boldsymbol{\hat{z}}}{ r}\bigg)^2\Bigg) + V_{pp\sigma}(r) \bigg(\frac{\boldsymbol{r}\cdot \boldsymbol{\hat{z}}}{ r}\bigg)^2, \nonumber \\
    &V_{pp\pi}(r) = V_{pp \pi}^0 e^{-(r - a_0)/r_0}, \nonumber \\ 
    &V_{pp\sigma}(r) = V_{pp \sigma}^0 e^{-(r - d_0)/r_0}, 
\end{align}
with the parameters $V^0_{ppp\pi} = 2.7$ eV,  $V^0_{pp\sigma} = 0.48$ eV and $r_0 = 0.0453 $ nm. 

\section{The symmetry operations under magnetic fields}\label{appb}

We look for unitary operators realizing the $C_{3z}$ and $C_{2z}$ symmetries, acting on the creation operators as
\begin{align}
g c^\dagger_{\boldsymbol{i}} g^{-1} = \exp(i\chi_g(g(\boldsymbol{r_i}))) c^\dagger_{g(\boldsymbol{i})}.
\end{align}
Here we use indistinctly $g$ for the unitary operators and for the linear transformations acting on points of the lattice. These can always be distinguished by the context.
% As in the main text, $c^\dagger_{\boldsymbol{i}}(c^\dagger_{g(\boldsymbol{i})})$ is the creation operator at position $\boldsymbol{r_i}$ $(g(\boldsymbol{r_i}))$.
The action on the Hamiltonian is
% \begin{align}
% \end{align}
\begin{align}
g H_0 g^{-1} = \sum_{\boldsymbol{i},\boldsymbol{j}} t(\boldsymbol{r_i}-\boldsymbol{r_j}) \exp(i\theta_{\boldsymbol{i},\boldsymbol{j}})\exp(i\chi_g(g(\boldsymbol{r_i})) - i\chi_g(g(\boldsymbol{r_j}))) c_{g(\boldsymbol{i})}^\dagger c_{g(\boldsymbol{j})}.
\end{align}

We are dealing with symmetries at zero flux, so $t(g(\boldsymbol{r_i})-g(\boldsymbol{r_j})) = t(\boldsymbol{r_i} - \boldsymbol{r_j})$. Then to realize the symmetry, i.e. for $g H_0 g^{-1} = H_0$, $\chi_g(\boldsymbol{r})$ must obey 
\begin{align}
\theta_{g^{-1}(\boldsymbol{i}), g^{-1}(\boldsymbol{j})} + \chi_g(\boldsymbol{r_i}) - \chi_g(\boldsymbol{j}) &= \theta_{\boldsymbol{i},\boldsymbol{j}} \nonumber \\
\frac{2\pi}{\Phi_0}\int_{g^{-1}(\boldsymbol{r_i})}^{g^{-1}(\boldsymbol{r_j})} \boldsymbol{A}(\boldsymbol{r'}) \cdot d\boldsymbol{r'} - \frac{2\pi}{\Phi_0}\int_{\boldsymbol{r_i}}^{\boldsymbol{r_j}} \boldsymbol{A}(\boldsymbol{r'}) \cdot d\boldsymbol{r'} &= \int_{\boldsymbol{r_i}}^{\boldsymbol{r_j}} \boldsymbol{\nabla} \chi_g(\boldsymbol{r'}) \cdot d\boldsymbol{r'} \nonumber \\
\frac{2\pi}{\Phi_0}\Bigg(g\big(\boldsymbol{A}(g^{-1}(\boldsymbol{r}))\big) - \boldsymbol{A}(\boldsymbol{r})\Bigg) &= \boldsymbol{\nabla}\chi_g(\boldsymbol{r}).
\end{align}
In the periodic Landau gauge, $\boldsymbol{A}(\boldsymbol{r}) = \frac{\Phi}{2\pi}\Big(\xi_1 \boldsymbol{G}_2 - 2\pi\boldsymbol{\nabla}\big(\xi_2\left \lfloor{\xi_1 + \epsilon}\right \rfloor\big) \Big) $, where $\xi_{1}$ and $\xi_{2}$ are defined by $\boldsymbol{r} = \xi_{1} \boldsymbol{L}_1 + \xi_{2} \boldsymbol{L}_2$. We have for $C_{3z}$
\begin{align}
C_{3z}\big(\boldsymbol{A}(C_{3z}^{-1}(\boldsymbol{r}))\big) = \frac{\Phi}{2\pi}\Bigg(-\xi_2 (\boldsymbol{G}_2 - \boldsymbol{G}_1) + 2\pi \boldsymbol{\nabla}\bigg((\xi_2+ \xi_1)\left \lfloor{\xi_2 + \epsilon}\right \rfloor\bigg) \Bigg), 
\end{align}
and hence
\begin{align}
\chi_{C_{3z}}(\boldsymbol{r}) = \frac{2 \pi p}{q} \Bigg(( \xi_1 + \xi_2)\lfloor \xi_2 + \epsilon \rfloor + \xi_2 \lfloor \xi_1 + \epsilon  \rfloor - \xi_1 \xi_2 - \frac{\xi_2^2}{2}\Bigg).
\end{align}
Similarly for $C_{2z}$ we get
\begin{align}
C_{2z}\Big(\boldsymbol{A}\big(C_{2z}^{-1}(\boldsymbol{r})\big)\Big) &= \frac{\Phi}{2\pi}\Bigg(\xi_1 \boldsymbol{G}_2  + 2\pi \boldsymbol{\nabla}\Big(\xi_2\left \lfloor{-\xi_1 + \epsilon}\right \rfloor\Big) \Bigg),
\end{align}
and hence
\begin{align}
\chi_{C_{2z}}(\boldsymbol{r}) &= \frac{2\pi p}{q} \Bigg(\xi_2 \lfloor \xi_1 + \epsilon \rfloor + \xi_2 \lfloor - \xi_1 + \epsilon \rfloor \Bigg). 
\end{align}
Above we have used the facts that for orthogonal transformations $g$ and scalar functions $f(\boldsymbol{r})$ and $h(\boldsymbol{r}) = f(g^{-1}(\boldsymbol{r}))$, we have $\boldsymbol{\nabla} h|_{\boldsymbol{r}}  = g\big(\boldsymbol{\nabla}f|_{g^{-1}(\boldsymbol{r})}\big)$, and that for a function of $\xi_1$ and $\xi_2$ we have $2\pi \boldsymbol{\nabla} f = \frac{\partial f}{\partial \xi_1} \boldsymbol{G}_1+ \frac{\partial f}{\partial \xi_2}  \boldsymbol{G}_2$. The functions $\chi_{C_{3z}}$ and $\chi_{C_{2z}}$ have the following periodicity properties 
\begin{align}
    \chi_{C_{3z}}(\boldsymbol{r} + q \boldsymbol{L}_2) =  \chi_{C_{3z}}(\boldsymbol{r}) + \pi p q \  \text{mod} \ 2\pi,& \quad  \chi_{C_{3z}}(\boldsymbol{r} + \boldsymbol{L}_1) = \chi_{C_{3z}}(\boldsymbol{r}) + \frac{2 \pi p}{q}\lfloor \xi_2 + \epsilon \rfloor \nonumber \\
    \chi_{C_{2z}}(\boldsymbol{r} + q \boldsymbol{L}_2) =  \chi_{C_{2z}}(\boldsymbol{r}) \  \text{mod} \ 2\pi,& \quad  \chi_{C_{2z}}(\boldsymbol{r} + \boldsymbol{L}_1) = \chi_{C_{2z}}(\boldsymbol{r}).
\end{align}
We are interested in $p=q=1$, so we can write
\begin{align}
e^{i \chi_{C_{3z}}(\boldsymbol{r})} =& e^{-i\boldsymbol{G}_2\cdot \boldsymbol{r}/2}e^{i\overline{\chi}_{C_{3z}}(\boldsymbol{r})} \nonumber \\
e^{i \chi_{C_{2
z}}(\boldsymbol{r})} = & e^{i \overline{\chi}_{C_{2z}}(\boldsymbol{r})},
\end{align}
where barred phases are periodic in the Moiré unit cell. As we will see now, the phases $e^{i \chi_{C_{3z}}(\boldsymbol{r})}$ and $e^{i \chi_{C_{2z}}(\boldsymbol{r})}$ modify the transformations of the Bloch waves, redefining the high symmetry points in flux.

The Bloch waves are written
\begin{align}
    c^\dagger_{\boldsymbol{k},\boldsymbol{i}} = \frac{1}{\sqrt{N_M}} \sum_{\boldsymbol{l}} e^{i\boldsymbol{k}\cdot (\boldsymbol{R_l} + \boldsymbol{\delta_i})} c^\dagger_{\boldsymbol{l},\boldsymbol{i}},
\end{align}
with $\boldsymbol{k}$ belonging to the Moiré Brillouin zone, and here $c^\dagger_{\boldsymbol{l},\boldsymbol{i}}$ creates an electron at position $\boldsymbol{R_l} + \boldsymbol{\delta_i}$ where $\boldsymbol{R_l}$ is a lattice vector and $\boldsymbol{\delta_i}$ belongs to the Wigner-Seitz cell. Under $C_{3z}$, $  c^\dagger_{\boldsymbol{k},\boldsymbol{i}}$ transforms as
\begin{align}
    C_{3z} c^\dagger_{\boldsymbol{k},\boldsymbol{i}} (C_{3z})^{-1} = \frac{1}{\sqrt{N_M}} \sum_{\boldsymbol{l}} e^{i(C_{3z}(\boldsymbol{k}) - \boldsymbol{G}_2/2)\cdot  C_{3z}(\boldsymbol{R_l} + \boldsymbol{\delta_i})} e^{i\overline{\chi}_{C_{3z}}(C_{3z}(\boldsymbol{\delta_i}))}c^\dagger_{C_{3z}(\boldsymbol{l},\boldsymbol{i})}.
\end{align}
Here, $c^\dagger_{C_{3z}(\boldsymbol{l},\boldsymbol{i})}$ creates an electron at position $C_{3z}(\boldsymbol{R_l} + \boldsymbol{\delta_i})$. We see that $C_{3z}$ sends momentum $\boldsymbol{k}$ to $C_{3z}(\boldsymbol{k}) - \boldsymbol{G}_2/2$. Via the embedding relation $c^\dagger_{\boldsymbol{k} + \boldsymbol{G},\boldsymbol{i}} = e^{i\boldsymbol{G}\cdot \boldsymbol{\delta_i}} c^\dagger_{\boldsymbol{k},\boldsymbol{i}}$ for $\boldsymbol{G}$ a reciprocal lattice vector, the three-fold rotation in flux acts in the momenta as follows,
\begin{align}
    \boldsymbol{k} \xrightarrow{C_{3z}} C_{3z}(\boldsymbol{k}) - \boldsymbol{G}_2/2 \sim C_{3z}\Big(\boldsymbol{k} - (\boldsymbol{G}_1 + \boldsymbol{G}_2)/2\Big) + (\boldsymbol{G}_1 + \boldsymbol{G}_2)/2.
\end{align}
Also, given that $\chi_{C_{2z}}(\boldsymbol{r})$ is periodic mod $2\pi$ on the unit cell, the momentum transforms like in zero flux, 
\begin{align}
    \boldsymbol{k} \xrightarrow{C_{2z}} C_{2z}(\boldsymbol{k}) \sim C_{2z}\Big(\boldsymbol{k} - (\boldsymbol{G}_1 + \boldsymbol{G}_2)/2\Big) + (\boldsymbol{G}_1 + \boldsymbol{G}_2)/2.
\end{align}
The center of rotations has shifted from $\Gamma_M = \boldsymbol{0}$ to $(\boldsymbol{G}_1 + \boldsymbol{G}_2)/2$ at one magnetic flux quantum.

Now we look for the operator realizing $C_{2y}$. The procedure is the same, but in this case $C_{2y}H_0(C_{2y})^{-1}$ should be equal to $H_0$ but with the sign of the magnetic field reversed. Hence, $\chi_{C_{2y}}(\boldsymbol{r})$ must obey
\begin{align}
    \frac{2\pi}{\Phi_0}\Bigg(C_{2y}\big(\boldsymbol{A}(C_{2y}^{-1}(\boldsymbol{r}))\big) + \boldsymbol{A}(\boldsymbol{r})\Bigg) = \boldsymbol{\nabla}\chi_{C_{2y}}(\boldsymbol{r}).
\end{align}
We obtain for $\chi_{C_{2y}}(\boldsymbol{r})$
\begin{align}
    \chi_{C_{2y}}(\boldsymbol{r}) = \frac{2 \pi p}{q} \Bigg(-\xi_2\lfloor \xi_1 + \epsilon \rfloor + \xi_2 \lfloor \xi_1 + \xi_2 + \epsilon  \rfloor - \frac{\xi_2^2}{2}\Bigg),
\end{align} which obeys the properties
\begin{align}
     \chi_{C_{2y}}(\boldsymbol{r} + q \boldsymbol{L}_2) =  \chi_{C_{2y}}(\boldsymbol{r}) - \pi p q \  \text{mod} \ 2\pi,& \quad  \chi_{C_{2y}}(\boldsymbol{r} + \boldsymbol{L}_1) = \chi_{C_{2y}}(\boldsymbol{r}).
\end{align}
Proceeding similarly to above, we get that under $C_{2y}$ the momentum transform as
\begin{align}
    \boldsymbol{k} \xrightarrow{C_{2y}} C_{2y}(\boldsymbol{k}) - \boldsymbol{G}_2/2 \sim C_{2y}\Big(\boldsymbol{k} - (\boldsymbol{G}_1 + \boldsymbol{G}_2)/2\Big) +  (\boldsymbol{G}_1 + \boldsymbol{G}_2)/2.
\end{align}

For the time reversal operator $\mathcal{T}$, the magnetic field should be reversed also, and it is trivial to see that the action is the same as for the zero flux case. $\mathcal{T}$ is an antiunitary operator satisfying $\mathcal{T}c^\dagger_{\boldsymbol{i}}\mathcal{T}^{-1} = c_{\boldsymbol{i}}^\dagger$, and transforming the momentum as 
\begin{align}
    \boldsymbol{k} \xrightarrow{\mathcal{T}} - \boldsymbol{k} \sim -\Big(\boldsymbol{k} - (\boldsymbol{G}_1 + \boldsymbol{G}_2)/2\Big) + (\boldsymbol{G}_1 + \boldsymbol{G}_2)/2.
\end{align}

It is important to notice here that when considering the combined operators $C_{2y}^2=\mathcal{T}^2=1, C_{2y}\mathcal{T}, \mathcal{T} C_{2y}$, the second operator acts on the system with the reversed magnetic flux because the first application changes the sign of the field. As a consequence, when $C_{2y}$ is the last operator, one must be reverse the $C_{2y}$ phase, $\chi_{C_{2y}} \to -\chi_{C_{2y}}$.

In conclusion, the action of symmetry operators under one magnetic flux quantum effectively shift the Brillouin zone by $(\boldsymbol{G}_1 + \boldsymbol{G_2})/2$, redefining the high symmetry points to $\Gamma = (\boldsymbol{G}_1 + \boldsymbol{G}_2)/2$, $M_M = \boldsymbol{G}_1/2$ $K_M=(1/6)\boldsymbol{G}_1 + (5/6)\boldsymbol{G_2}$ and $K'_M=(5/6)\boldsymbol{G}_1 + (1/6)\boldsymbol{G_2}$.

\section{Valley charge and $C_{2z}P$ operator on the lattice}\label{appc}

We wish to find an operator $\tau_z$ implementing the valley charge on the lattice, such that $\langle \tau_z \rangle = +1$ on states nearby the $K$ point of graphene and $-1$ near the $K'$ point. We adopt a slight generalization of the valley operator of Ref. \cite{Lado19}
% \begin{align}
%     \tau_z = \sum_{\triangle} e^{i \theta_{\boldsymbol{}}}c^\dagger_{\boldsymbol{r}(\triangle(1))}  c^\dagger_{\boldsymbol{r}(\triangle(2))} + c^\dagger_{\boldsymbol{r}(\triangle(2))}  c^\dagger_{\boldsymbol{r}(\triangle(3))} + c^\dagger_{\boldsymbol{r}(\triangle(3))}  c^\dagger_{\boldsymbol{r}(\triangle(1))}
% \end{align}
\begin{align}
    \tau_z = \frac{i}{3\sqrt{3}}\sum_{l}\bigg(&\sum_{\bigtriangledown} e^{-i \theta_{\bigtriangledown(1),\bigtriangledown(2)}} c^\dagger_{\bigtriangledown(1)}  c_{\bigtriangledown(2)} + e^{-i \theta_{\bigtriangledown(2),\bigtriangledown(3)}} c^\dagger_{\bigtriangledown(2)}  c_{\bigtriangledown(3)} + e^{-i \theta_{\bigtriangledown(3),\bigtriangledown(1)}} c^\dagger_{\bigtriangledown(3)}  c_{\bigtriangledown(1)} \nonumber \\ &- \sum_{\triangle} e^{-i \theta_{\triangle(1),\triangle(2)}} c^\dagger_{\triangle(1)}  c_{\triangle(2)} + e^{-i \theta_{\triangle(2),\triangle(3)}} c^\dagger_{\triangle(2)}  c_{\triangle(3)} + e^{-i \theta_{\triangle(3),\triangle(1)}} c^\dagger_{\triangle(3)}  c_{\triangle(1)}  \bigg) + \text{h.c.}.
\end{align}

\begin{figure}[H]
  \hspace{-2.5cm} $a)$ \hspace{5.5cm}   $b)$ \\
  \centering
  \includegraphics[width=.32\linewidth]{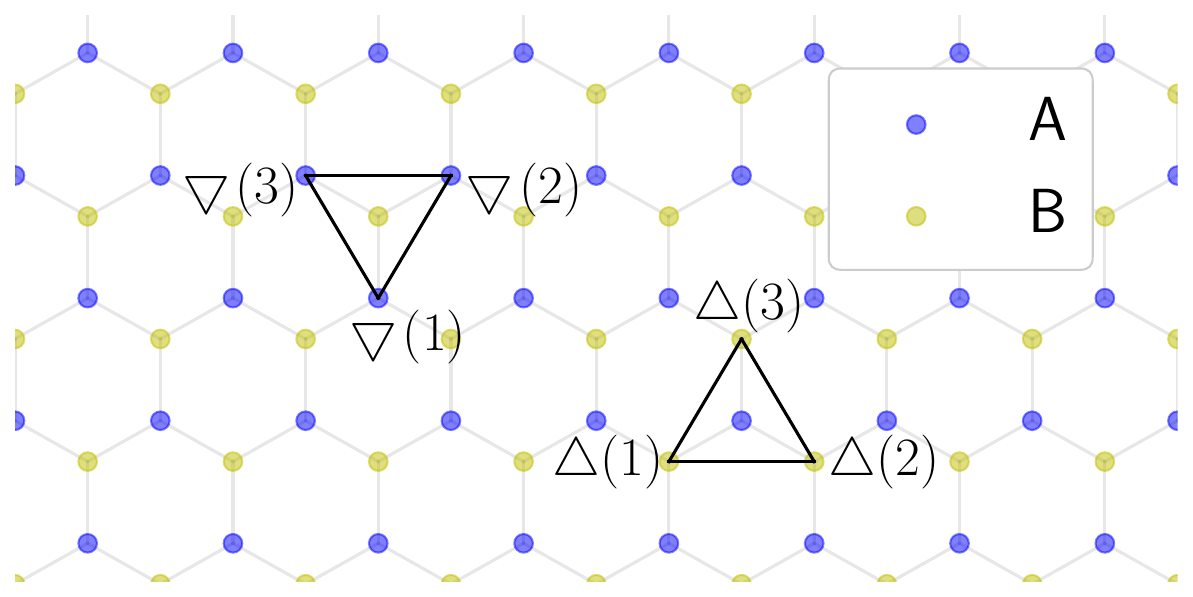}
  \includegraphics[width=.2\linewidth]{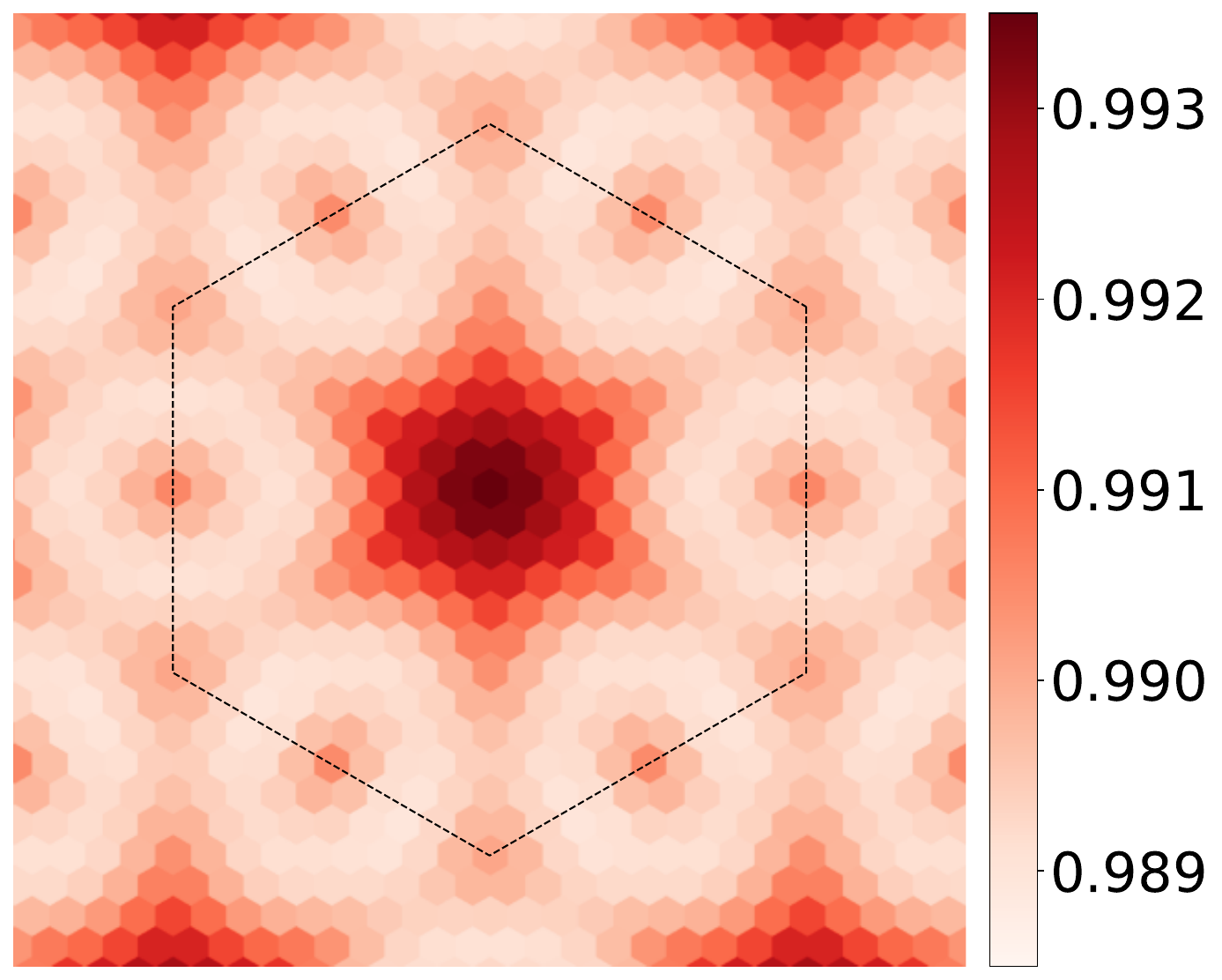}
  \includegraphics[width=.2\linewidth]{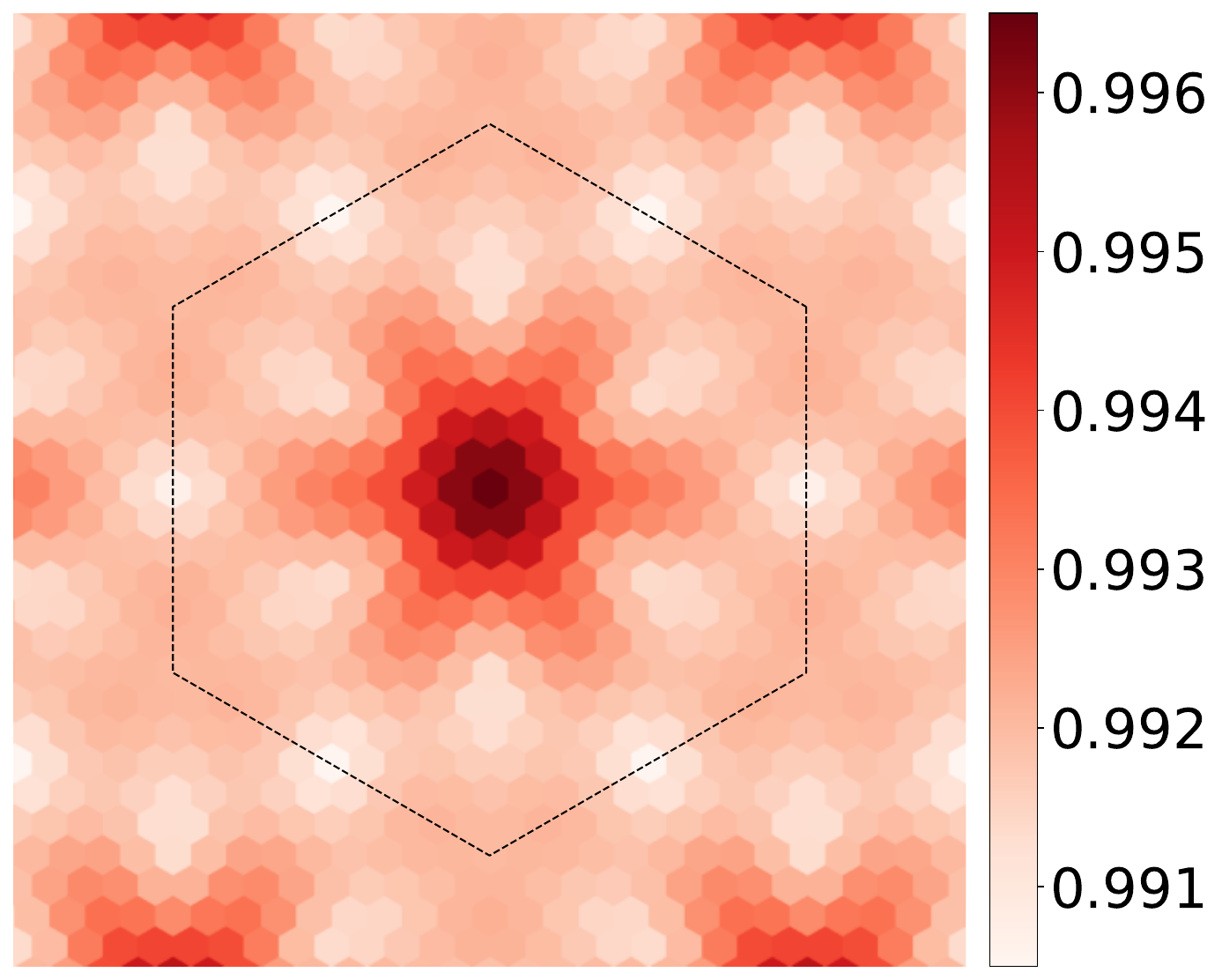}
\caption{\textbf{a)} Triangular loops that compute the valley charge. \textbf{b)} Singular values of the $C_{2z}P$ operator projected onto the flat bands. The properties $\overline{C_{2z}P}(\boldsymbol{k})^\dagger = \overline{C_{2z}P}(\boldsymbol{k})$ and $\{\tau_z,\overline{C_{2z}P}(\boldsymbol{k})\}=0$ force them to be degenerate in pairs, so we show the two distinct ones. Unitary matrices have singular values equal to $1$.}
\label{valleyopsing}
\end{figure}

The sums are over triangles upside down of sublattice $A$ atoms, and triangles of sublattice $B$, and $l$ denotes the sum over the two layers. We draw an example of each kind of triangle in Figure \ref{valleyopsing}. The phases are the Peierls' phases defined in the main text. It can be shown that valley $K$ states have $\langle \tau_z \rangle = +1 + O(a/L_M)$ and valley $K'$  states have $\langle \tau_z \rangle = -1 + O(a/L_M)$. Diagonalization of the $\tau_z$ matrix in the flat bands $\langle \boldsymbol{k} \rho | \tau_z | \boldsymbol{k} \rho'\rangle$ outputs a valley polarized basis.

On the other hand, a general wave function can be written in first quantized notation (here we omit the spin)
\begin{align}
    |\psi\rangle = \sum_{\boldsymbol{r_i}}\psi(\boldsymbol{r_i}) |\boldsymbol{r_i}\rangle = \sum_{\eta \sigma l} \sum_{\boldsymbol{r_i}\in \sigma l} e^{i \eta \boldsymbol{K_l} \cdot \boldsymbol{r_i}} f^\psi_{\eta \sigma l}(\boldsymbol{r_i})  |\boldsymbol{r_i}\rangle
\end{align}
where the $f$ envelopes depend on layer $l=$ top(t), botttom(b), and sublattice $\sigma=A,B$, and the valley phases are rapidly oscillating. $\boldsymbol{K_t} = R_{-\theta/2}(-4\pi/3a,0)$ and $\boldsymbol{K_b}=R_{\theta/2}(-4\pi/3a,0)$ are depicted in Fig. \ref{bandsnonint}a) of the main text. In the continuum model, the $f$ functions are promoted to smooth functions of $\boldsymbol{r}$.

% , and $\Psi$ can be written as a vector,
% \begin{align}
%     \Psi(\boldsymbol{r}) = \Big(f_{KAt}(\boldsymbol{r}), f_{KBt}(\boldsymbol{r}), f_{K'At}(\boldsymbol{r}), f_{K'Bt}(\boldsymbol{r}), f_{KAb}(\boldsymbol{r}), f_{KBb}(\boldsymbol{r}), f_{K'Ab}(\boldsymbol{r}), f_{K'Bb}(\boldsymbol{r})\Big).
% \end{align}
The particle-hole operator $C_{2z}P$ is defined in the continuum wave functions, interchanging the valley, sublattice and layer,
% \begin{align}
%     C_{2z}P (\Psi) (\boldsymbol{r}) =  \Big(f_{K'Bb}(\boldsymbol{r}), f_{K'Ab}(\boldsymbol{r}), -f_{KBb}(\boldsymbol{r}), -f_{KAb}(\boldsymbol{r}), -f_{K'Bt}(\boldsymbol{r}), -f_{K'At}(\boldsymbol{r}), f_{KBt}(\boldsymbol{r}), f_{KAt}(\boldsymbol{r})\Big).
% \end{align}
\begin{align}
    f^{C_{2z}P(\psi)}_{\eta\sigma l}(\boldsymbol{r}) = \eta s_l f^\psi_{-\eta \Bar{\sigma}\Bar{l}}(\boldsymbol{r}),
\end{align}
with $s_l= 1(-1)$ for $l=t(b)$ and $\Bar{\sigma}$ and $\Bar{l}$ denote the opposite sublattice and layer to $\sigma$ and $l$. Notice that it is a local operator, so it will not change the momentum of a Bloch state. 

\begin{figure}[H]
\centering
  \centering
  \includegraphics[width=.37\linewidth]{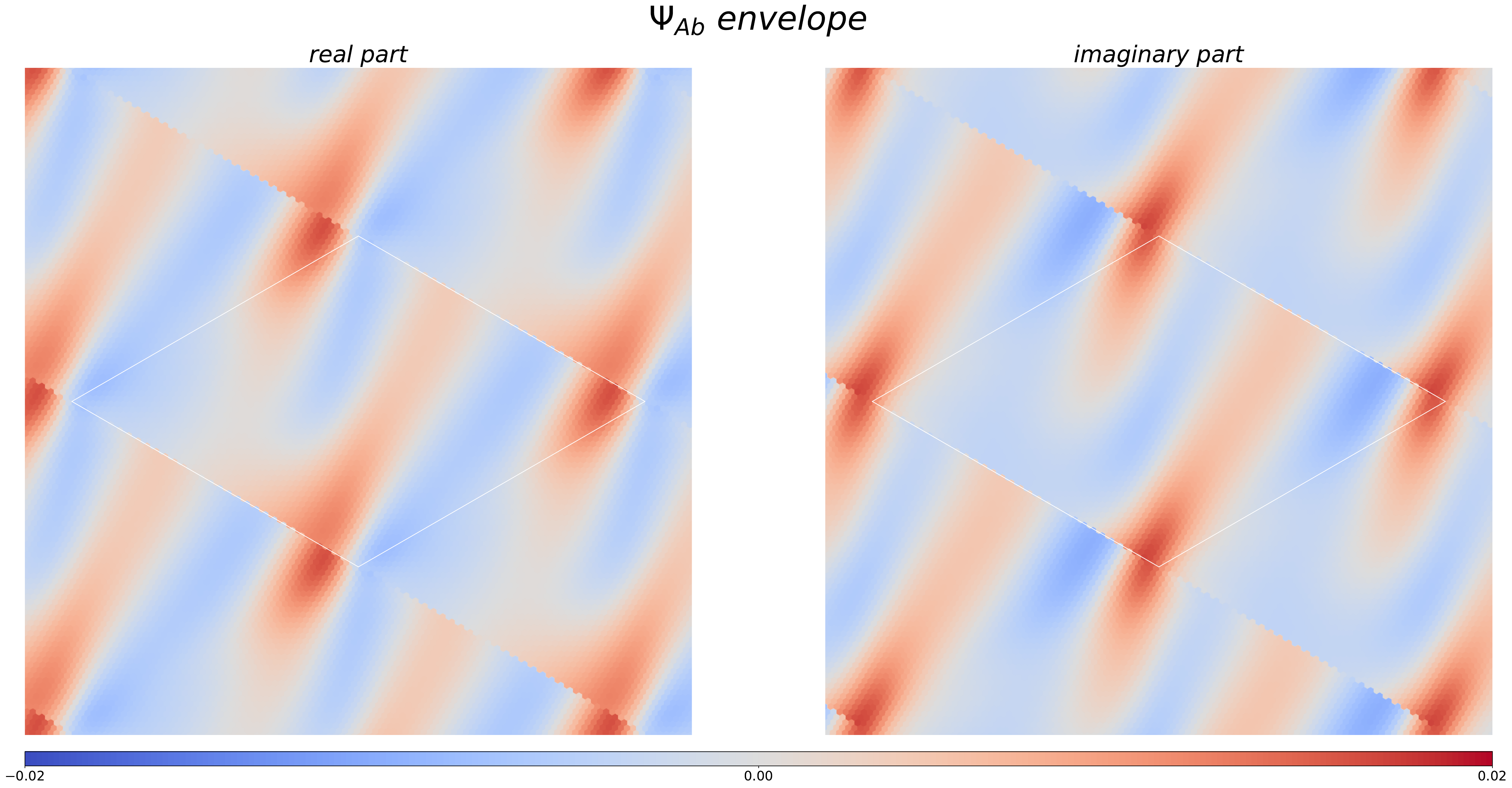}
  \includegraphics[width=.37\linewidth]{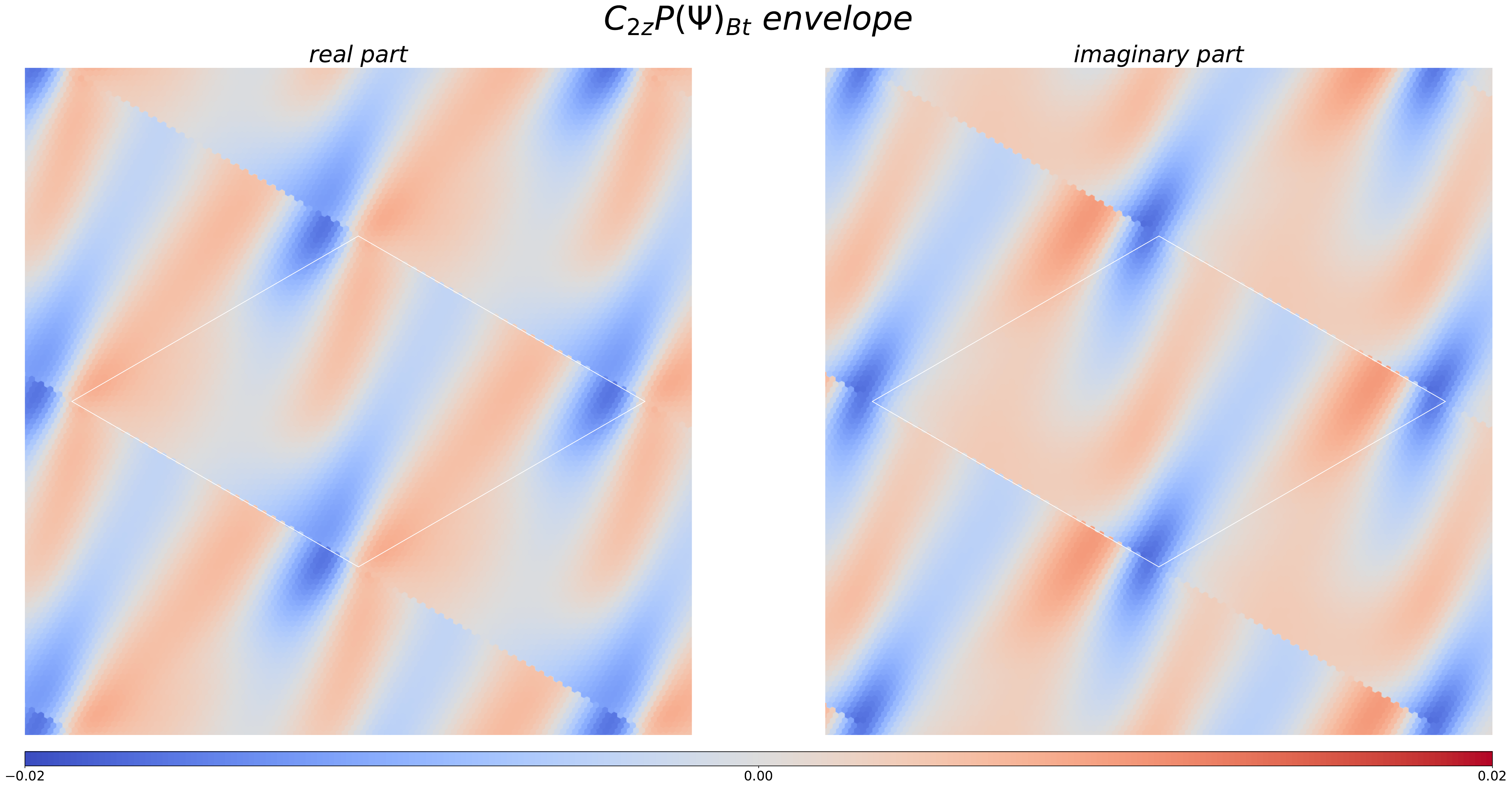}
\caption{The envelope function of valley $K$, sublattice $A$, bottom layer of a Bloch state to the left, and the envelope function of valley $K'$, sublattice $B$, top layer of the $C_{2z}P$ transformed state to the right. Notice the discontinuity at the cell edge due to the periodic Landau gauge.}
\label{valleyopc2p}
\end{figure}

On the lattice, $C_{2z}P$ has to be effectively defined as follows. In a valley polarized basis, we obtain the envelope functions by removing the corresponding valley phases. Afterwards, we perform a smooth interpolation of the data $f_{\eta \sigma l}(\boldsymbol{r_i})$, being $\boldsymbol{r_i}$ the positions of the atoms at sublattice $\sigma$ and layer $l$. Finally, the smooth functions are sampled at the points of the opposite sublattice and layer and the new valley phase is incorporated. In Fig. \ref{valleyopc2p} we show an example of the envelope functions before and after this procedure. As a note, the envelope functions have a discontinuity at $\xi_1 = $ integer in the periodic Landau gauge, and some care is needed when performing the interpolations.

The projected operator in the flat bands $[\overline{C_{2z}P}(\boldsymbol{k})]_{\rho \rho'} = \langle \boldsymbol{k} \rho | C_{2z}P | \boldsymbol{k}\rho' \rangle$ is then constructed in the basis of choice. We have checked that the particular basis is irrelevant, and the matrix elements of $\overline{C_{2z}P}(\boldsymbol{k})$ in a new basis computed via unitary conjugation of the first and via interpolation in the new basis are essentially identical. We also checked that the properties $\overline{C_{2z}P}(\boldsymbol{k})^\dagger = \overline{C_{2z}P}(\boldsymbol{k})$ and $\{\tau_z,\overline{C_{2z}P}(\boldsymbol{k})\}=0$ are preserved by our procedure, with matrix elements of the $\tau_z$-commuting or anti-hermitian parts always less than $10^{-5}$.

\section{Symmetry of the model}\label{appd}

% A Bloch state $|\psi \rangle$ of valley $\eta$ has a wave function, in first quantized notation,
% \begin{align}
%     | \psi \rangle = \sum_{r_{\boldsymbol{i}}} e^{i\eta \boldsymbol{K_l}\cdot \boldsymbol{r_i}}  f^\psi_{\eta \sigma_i l_i}(\boldsymbol{r_i}) |\boldsymbol{r_i} \rangle
% \end{align}
% with $f^\psi_{\eta \sigma l}(\boldsymbol{r})$ denoting graphene sublattice ($\sigma=A,B$) and layer ($l$) dependent smooth functions in the Moiré cell.

Consider a general matrix element of the Coulomb interaction between states $|i\rangle$ ($i=a,b,c,d$) with valleys $\eta_i$,
\begin{align}
    V_{abcd} = \frac{1}{2} \sum_{\sigma_i,l_i\atop{\sigma_j,l_j}}\sum_{\boldsymbol{r_i} \in \sigma_i l_i \atop{\boldsymbol{r_j} \in \sigma_j l_j}} e^{i(\eta_d - \eta_a)\boldsymbol{K_{l_i}}\cdot \boldsymbol{r_i} + (\eta_c - \eta_b)\boldsymbol{K_{l_j}}\cdot \boldsymbol{r_j}} V(\boldsymbol{r_i}-\boldsymbol{r_j}) f^a_{\eta_a \sigma_i l_i }(\boldsymbol{r_i})^* f^b_{\eta_b \sigma_j l_j }(\boldsymbol{r_j})^* f^c_{\eta_c \sigma_j l_j }(\boldsymbol{r_j}) f^d_{\eta_d \sigma_i l_i }(\boldsymbol{r_i}),
\end{align}
If we have something other than $\eta_a=\eta_d$ and $\eta_b=\eta_c$, then the rapidly oscillating phases will interfere in the sum over $\boldsymbol{r_{i,j}}$ and the matrix element vanishes. Putting in the spins $s_i$, we have
\begin{align}
    V_{abcd} \propto \delta_{\eta_a,\eta_d}\delta_{\eta_b,\eta_c} \delta_{s_a,s_d}\delta_{s_b,s_c}.
\end{align}
This general form of the interaction enjoys a $U(1)\times SU(2) \times SU(2)$ symmetry. The $U(1)$ is the valley charge conervation symmetry, acting as $|\boldsymbol{k}\eta \lambda \rangle \to e^{i\eta \phi}|\boldsymbol{k}\eta \lambda \rangle$, and the two $SU(2)$ correspond to independent spin rotations in each valley.

Furthermore, the states $|i'\rangle = C_{2z}P|i\rangle$ produce the matrix element
\begin{align}
    V_{a'b'c'd'} = \frac{1}{2}
   \sum_{\sigma_i,l_i\atop{\sigma_j,l_j}}\sum_{\boldsymbol{r_i} \in \sigma_i l_i \atop{\boldsymbol{r_j} \in \sigma_j l_j}} V(\boldsymbol{r_i}-\boldsymbol{r_j}) f^a_{\eta_a \Bar{\sigma}_i \Bar{l}_i }(\boldsymbol{r_i})^* f^b_{\eta_b \Bar{\sigma}_j \Bar{l}_j }(\boldsymbol{r_j})^* f^c_{\eta_c \Bar{\sigma}_j \Bar{l}_j }(\boldsymbol{r_j}) f^d_{\eta_d \Bar{\sigma}_i \Bar{l}_i }(\boldsymbol{r_i}),
\end{align}
where $\Bar{\sigma}$ and $\Bar{l}$ denote the opposite sublattice and layer to $\sigma$ and $l$. Replacing each $\boldsymbol{r_i}, \boldsymbol{r_j}$ by $\Bar{\boldsymbol{r}}_{\boldsymbol{i}}, \Bar{\boldsymbol{r}}_{\boldsymbol{j}}$ with approximately the same $x$ and $y$ coordinates (or less strictly, approximately the same $x$ and $y$ coordinates differences) but on opposite sublattices and layers, we get 
\begin{align}
    V_{a'b'c'd'} = \frac{1}{2}\sum_{\Bar{\sigma}_i,\Bar{l}_i\atop{\Bar{\sigma}_j,\Bar{l}_j}}\sum_{\boldsymbol{\Bar{r}_i} \in \Bar{\sigma}_i \Bar{l}_i \atop{\boldsymbol{\Bar{r}_j} \in \Bar{\sigma}_j \Bar{l}_j}} V(\Bar{\boldsymbol{r}}_{\boldsymbol{i}}-\Bar{\boldsymbol{r}}_{\boldsymbol{j}}) f^a_{\eta_a \Bar{\sigma}_i \Bar{l}_i }(\Bar{\boldsymbol{r}}_{\boldsymbol{i}})^* f^b_{\eta_b \Bar{\sigma}_j \Bar{l}_j }(\Bar{\boldsymbol{r}}_{\boldsymbol{j}})^* f^c_{\eta_c \Bar{\sigma}_j \Bar{l}_j }(\Bar{\boldsymbol{r}}_{\boldsymbol{j}}) f^d_{\eta_d \Bar{\sigma}_i \Bar{l}_i }(\Bar{\boldsymbol{r}}_{\boldsymbol{j}})
    =V_{abcd}.
    \label{vc2p}
\end{align}

We have established that $V_{abcd} = V_{a'b'c'd'}$. To conclude that the particle-hole operator generates a continuous symmetry we need $[C_{2z}P,V] = 0$, which is equivalent to
\begin{align}
    \sum_{abcd} V_{a'bbcd} - V_{abcd'} + V_{ab'cd} - V_{abc'd} = 0.
\end{align}
To show that $\sum_{abcd} V_{a'bcd} - V_{abcd'} = 0$, divide the basis vectors into two sets $S,S'$ such that $S'=C_{2z}P(S)$ and the union of $S$ and $S'$ is the complete basis. Then,
\begin{align}
    \sum_{abcd} V_{a'bcd} - V_{abcd'} =& \sum_{ad}\Bigg( \sum_{bc \in S} V_{a'bcd} +  \sum_{bc \in S'} V_{a'bcd} -  \sum_{bc \in S} V_{abcd'} -  \sum_{bc \in S'} V_{abcd'} \Bigg)  \nonumber \\
    =&\sum_{ad}\Bigg( \sum_{bc \in S} V_{a'bcd} -  \sum_{bc \in S'} V_{abcd'}  +  \sum_{bc \in S'} V_{a'bcd} -  \sum_{bc \in S} V_{abcd'}  \Bigg) \nonumber \\ =& \sum_{ad}\Bigg( \sum_{bc \in S} V_{a'bcd} -  \sum_{bc \in S} V_{a'bcd}  +  \sum_{bc \in S'} V_{a'bcd} -  \sum_{bc \in S'} V_{a'bcd}  \Bigg) = 0,
\end{align}
where we have used $V_{abcd} = V_{a'b'c'd'}$, $|a''\rangle = |a\rangle$ and $\sum_{a,b\in S'} O_{a'b'} = \sum_{a,b\in S} O_{ab}$. The identity $ \sum_{abcd}  V_{ab'cd} - V_{abc'd} = 0$ follows the the same way, and we conclude that $C_{2z}P$ generates another $U(1)$ symmetry of the Coulomb interaction. With the total charge conservation, the symmetry group is $U(4)$. It has $16$ generators $S_{ij}$ ($i,j=0,x,y,z$) with the following form in the irrep basis.
\begin{align}
    S_{ij} = \sum_{\boldsymbol{k}} [\lambda_0 \tau_i s_j]_{\rho \rho'}\ c^\dagger_{\boldsymbol{k}\rho} c^\dagger_{\boldsymbol{k}\rho'},
\end{align}
the index $\rho$ includes valley, irrep and spin, and $s_{0,x,y,z}$ denote the identity and Pauli matrices in spin space. 

In the total system, however, this large $U(4)$ symmetry is broken by several terms. First of all, the matrix elements with $\eta_a=-\eta_d=-\eta_b=\eta_c$ are small but nonzero, breaking the $SU(2) \times SU(2)$ down to the global $SU(2)$ of spin. One can show similarly to before that $H_U$ respects the $U(1)$ valley and $C_{2z}P$-generated symmetries, but breaks $SU(2) \times SU(2)$. This kind of valley-exchanging interactions have been termed intervalley Hund's couplings\cite{Bultinck20,Lee2019,Chatterjee2022}. 

Also, the kinetic energy breaks $C_{2z}P$ and the Zeeman energy preserves only the spin rotations around the $z$ axis. Finally, notice that there is an intrinsic breaking of $C_{2z}P$ due to the lattice (see the approximations we made to arrive to Eq. \ref{vc2p}) as well as and the flat-band projection (as discussed around Eq. \ref{c2pp1}). The $U(1)$ valley symmetry is preserved in the total system to a great accuracy.

\begin{figure}[H]
    \centering
    \includegraphics[width=.9\linewidth]{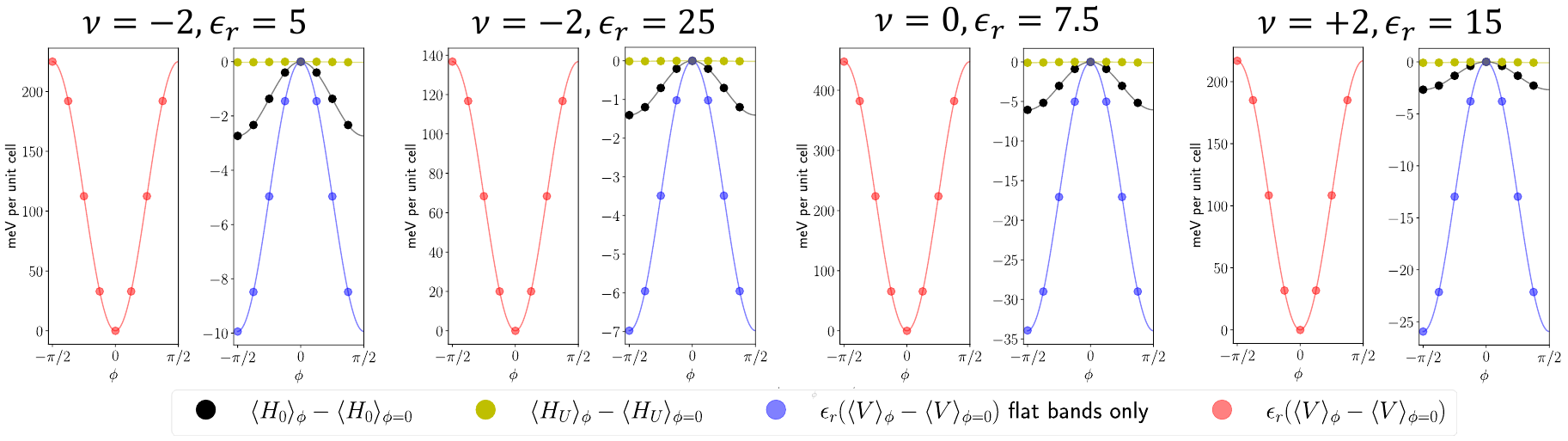}
    % \includegraphics[width=.45\linewidth]{figurec2pbr1.pdf}
    % \includegraphics[width=.45\linewidth]{figurec2pbr2.pdf}
    % \\
    % \includegraphics[width=.6\linewidth]{figurec2pbr3.pdf}
    \caption{\textbf{Kinetic, Hubbard and Coulomb energies of $\boldsymbol{\text{exp}(i\phi S_{x0})|\text{GS}\rangle}$} with respect to to the energies of the ground state $|\text{GS}\rangle$ for several selected parameters. Notice that the Coulomb energies have been multiplied by $\epsilon_r$. $U=4$ eV in all cases. Clearly, the approximate symmetry of the flat bands is broken by the Fermi sea electrons.}
    \label{c2pbreaking}
\end{figure}

All this contributions are comparatively small with respect to the symmetry-preserving part of the Coulomb energy, leading to the picture of the $U(4)$ ferromagnets in MATBG\cite{seo19,Kang19,Bultinck20,bernevig421,herzog22_3,singh2023topological,wang22}. However, the interactions with the Fermi sea break strongly the $U(1)$ subgroup generated by $C_{2z}P$. The electrons in the flat bands interact among themselves and with the mean field produced by the correlation matrix $\langle \text{FS} | c^\dagger_{\boldsymbol{i}s} c_{\boldsymbol{j}s} | \text{FS}  \rangle$ - $\langle 0 | c^\dagger_{\boldsymbol{i}s} c_{\boldsymbol{j}s} | 0  \rangle$, where $|\text{FS}\rangle$ is the state at $\nu=-4$ of occupied remote bands and $|0\rangle$ the reference state of the normal-order subtraction (see Appendix \ref{appe} for the details of the normal-ordering and flat-band projection). In Fig. \ref{c2pbreaking} we plot the energies of the states $\text{exp}(i\phi S_{x0})|\text{GS}\rangle$, corresponding to $C_{2z}P$ rotations of several selected ground states. We plot the total kinetic, Hubbard and Coulomb energies and the Coulomb energy restricted to the interactions of flat band electrons. Clearly, the kinetic, Hubbard and Coulomb flat-band physics are approximately symmetric, but the Fermi sea potential strongly breaks the symmetry.   

\section{The Hartree-Fock method and flat band projection}\label{appe}
Consider the normal ordered interaction of Eqs. \ref{potential} and \ref{hubbard},
\begin{align}
     V + H_U = \frac{1}{2}\sum_{\boldsymbol{r_i},\boldsymbol{r_j} s_i s_j} V(\boldsymbol{r_i}-\boldsymbol{r_j}) :c^\dagger_{\boldsymbol{i},s_i} c_{\boldsymbol{i}, s_i} c^\dagger_{\boldsymbol{j},s_j} c_{\boldsymbol{j}, s_j}: + U \sum_{\boldsymbol{i}} :c_{\boldsymbol{i}\uparrow}^\dagger c_{\boldsymbol{r_i}\uparrow} c_{\boldsymbol{i}\downarrow}^\dagger c_{\boldsymbol{i}\downarrow}:
\end{align}

The choice of the normal ordering with respect to the ground state of graphene at charge neutrality is necessary to avoid double counting the interaction\cite{xie20,Bultinck20}. This is, we assume that the hopping integrals $t(\boldsymbol{r})$ are already renormalized by the interactions with the deep Fermi sea of graphene. After expanding the normal ordered product\cite{giuliani_vignale_2005} and performing the Hartree-Fock decoupling, the Hamiltonian reads
\begin{align}
    V_{\text{HF}} + H_{U\text{HF}} =& \sum_{\boldsymbol{r_i},\boldsymbol{r_j},s_i,s_j} V(\boldsymbol{r_i}-\boldsymbol{r_j}) c_{\boldsymbol{i}s_i}^\dagger c_{\boldsymbol{i}s_i} \Big(\langle c_{\boldsymbol{j}s_j}^\dagger c_{\boldsymbol{j}s_j}\rangle - \langle c^\dagger_{\boldsymbol{j}s_j}c_{\boldsymbol{j}s_j}\rangle_0 \Big) -  \sum_{\boldsymbol{r_i},\boldsymbol{r_j},s} V(\boldsymbol{r_i}-\boldsymbol{r_j}) c_{\boldsymbol{i}s}^\dagger c_{\boldsymbol{j}s} \Big(\langle c_{\boldsymbol{i}s}^\dagger c_{\boldsymbol{j}s}\rangle - \langle c^\dagger_{\boldsymbol{i}s}c_{\boldsymbol{j}s}\rangle_0 \Big)^* \nonumber \\
    &+ U \sum_{\boldsymbol{r_i}} c_{\boldsymbol{i}\uparrow}^\dagger c_{\boldsymbol{i}\uparrow} \Big( \langle c_{\boldsymbol{i}\downarrow}^\dagger c_{\boldsymbol{i}\downarrow} \rangle - \langle c_{\boldsymbol{i}\downarrow}^\dagger c_{\boldsymbol{i}\downarrow} \rangle_0 \Big) + U \sum_{\boldsymbol{r_i}} c_{\boldsymbol{i}\downarrow}^\dagger c_{\boldsymbol{i}\downarrow} \Big( \langle c_{\boldsymbol{i}\uparrow}^\dagger c_{\boldsymbol{i}\uparrow} \rangle - \langle c_{\boldsymbol{i}\uparrow}^\dagger c_{\boldsymbol{i}\uparrow} \rangle_0 \Big) + \text{constant},
\end{align}
with $\langle ... \rangle_0$ denoting the expectation value in the ground state of graphene at charge neutrality, and $\langle ... \rangle$ the expectation value in the particular state of our Hartree-Fock decoupling. In our implementation we restrict the wave function to be a direct product of spin up and down electrons, such that $\langle c^\dagger_{\boldsymbol{i}\uparrow} c_{\boldsymbol{j}\downarrow} \rangle = 0$ for all $\boldsymbol{r_i},\boldsymbol{r_j}$.

In the projected limit we assume that the remote bands are filled and the relevant physics takes place in the flat bands. In this spirit we compute mean field interaction restricted to the subspace of the flat bands,
\begin{align}
    [V_{\text{HF,p}}(\boldsymbol{k},\boldsymbol{k'}) + H_{U\text{HF,p}}(\boldsymbol{k},\boldsymbol{k'})]_{\rho \rho'} = \Big(\langle \text{FS}| \otimes \langle \boldsymbol{k}\rho | \Big) \Big( V_{\text{HF}} + H_{U\text{HF}}\Big) \Big(| \text{FS} \rangle \otimes | \boldsymbol{k'} \rho' \rangle \Big),
\end{align}
with $ |\text{FS}\rangle \otimes |\boldsymbol{k}\rho \rangle$ denoting the direct product of the state with the filled remote bands and the state with momentum $\boldsymbol{k}$ and multi-index $\rho$. We further assume translational symmetry that makes the mean field Hamiltonian block-diagonal in momentum space, $V_{\text{HF,p}}(\boldsymbol{k},\boldsymbol{k'}) + H_{U\text{HF,p}}(\boldsymbol{k},\boldsymbol{k'}) = \Big(V_{\text{HF,p}}(\boldsymbol{k}) + H_{U\text{HF,p}}(\boldsymbol{k})\Big)\delta_{\boldsymbol{k},\boldsymbol{k'}}$.

The self-consistent method starts by proposing an ansatz for the ground state at any given filling, computing the mean field Hamiltonian and performing the flat band projection. Next, we solve the projected mean filed Hamiltonian 
\begin{align}
H_{\text{HF,p}}(\boldsymbol{k},\boldsymbol{k'}) = \Big(H_{0\text{,p}}(\boldsymbol{k}) + V_{\text{HF,p}}(\boldsymbol{k})+H_{U\text{HF,p}}(\boldsymbol{k})\Big) \delta_{\boldsymbol{k},\boldsymbol{k'}},    
\end{align}
with $H_{0\text{,p}}(\boldsymbol{k})\delta_{\boldsymbol{k},\boldsymbol{k'}}$ the projected kinetic energy operator. The ground state of this Hamiltonian is then a new ansatz for the self-consistent ground state and the process is repeated until convergence is reached.

The energy of the self-consistent state is
\begin{align}
    \langle H \rangle =& \langle V \rangle + \langle H_U \rangle + \langle H_0 \rangle \nonumber \\
    =& \frac{1}{2} \sum_{\boldsymbol{r_i},\boldsymbol{r_j},s_i,s_j} V(\boldsymbol{r_i}-\boldsymbol{r_j}) \Big(\langle c_{\boldsymbol{i}s_i}^\dagger c_{\boldsymbol{i}s_i}\rangle - \langle c^\dagger_{\boldsymbol{i}s_i}c_{\boldsymbol{i}s_i}\rangle_0 \Big) \Big(\langle c_{\boldsymbol{j}s_j}^\dagger c_{\boldsymbol{j}s_j}\rangle - \langle c^\dagger_{\boldsymbol{j}s_j}c_{\boldsymbol{j}s_j}\rangle_0 \Big) - \frac{1}{2} \sum_{\boldsymbol{r_i},\boldsymbol{r_j},s} V(\boldsymbol{r_i}-\boldsymbol{r_j}) \Big\lvert \Big\lvert\langle c_{\boldsymbol{i}s}^\dagger c_{\boldsymbol{j}s}\rangle - \langle c^\dagger_{\boldsymbol{i}s}c_{\boldsymbol{j}s}\rangle_0 \Big\rvert \Big\rvert^2 \nonumber \\ &+ U \sum_{\boldsymbol{r_i}} \Big(\langle c_{\boldsymbol{i}\uparrow}^\dagger c_{\boldsymbol{i}\uparrow}\rangle - \langle c^\dagger_{\boldsymbol{i}\uparrow}c_{\boldsymbol{i}\uparrow}\rangle_0 \Big) \Big(\langle c_{\boldsymbol{i}\downarrow}^\dagger c_{\boldsymbol{i}\downarrow}\rangle - \langle c^\dagger_{\boldsymbol{i}\downarrow}c_{\boldsymbol{i}\downarrow}\rangle_0 \Big) + \sum_{\boldsymbol{r_i},\boldsymbol{r_j}s} t(\boldsymbol{r_i}-\boldsymbol{r_j})e^{i\theta_{\boldsymbol{i},\boldsymbol{j}}} \langle c_{\boldsymbol{i}s}^\dagger c_{\boldsymbol{j}s} \rangle.
\end{align}
The Coulomb interaction is split into the Hartree or direct and Fock or exchange terms, with the plus and minus signs in front respectively. 
% In our algorithm, we always work with the Fock matrix $\langle c_{\boldsymbol{i}s}^\dagger c_{\boldsymbol{i}s}\rangle - \langle c^\dagger_{\boldsymbol{i}s}c_{\boldsymbol{i}s}\rangle_0$, so the values reported for the kinetic energy have a constant offset of $\sum_{\boldsymbol{r_i},\boldsymbol{r_j}s} t(\boldsymbol{r_i}-\boldsymbol{r_j})e^{i\theta_{\boldsymbol{i},\boldsymbol{j}}} \langle c_{\boldsymbol{i}s}^\dagger c_{\boldsymbol{j}s} \rangle_0$.
% It is important to notice that both the Hartee and Fock energies, as they are written above, have contributions which are not physical coming from self interactions of the electron. These contributions couple the density of one of the orbitals in the Slater determinant with itself. They have opposite signs in the Hartree and Fock parts and cancel each other in the total energy.

\section{Additional plots of the Hartree-Fock simulations}\label{appf}

\begin{figure}[H]
  \hspace{-3cm} $a)$ \hspace{4.7cm} $b)$ \hspace{7.7cm} $c)$ \hspace{20cm} \\
    \centering
    \includegraphics[width=.13\linewidth]{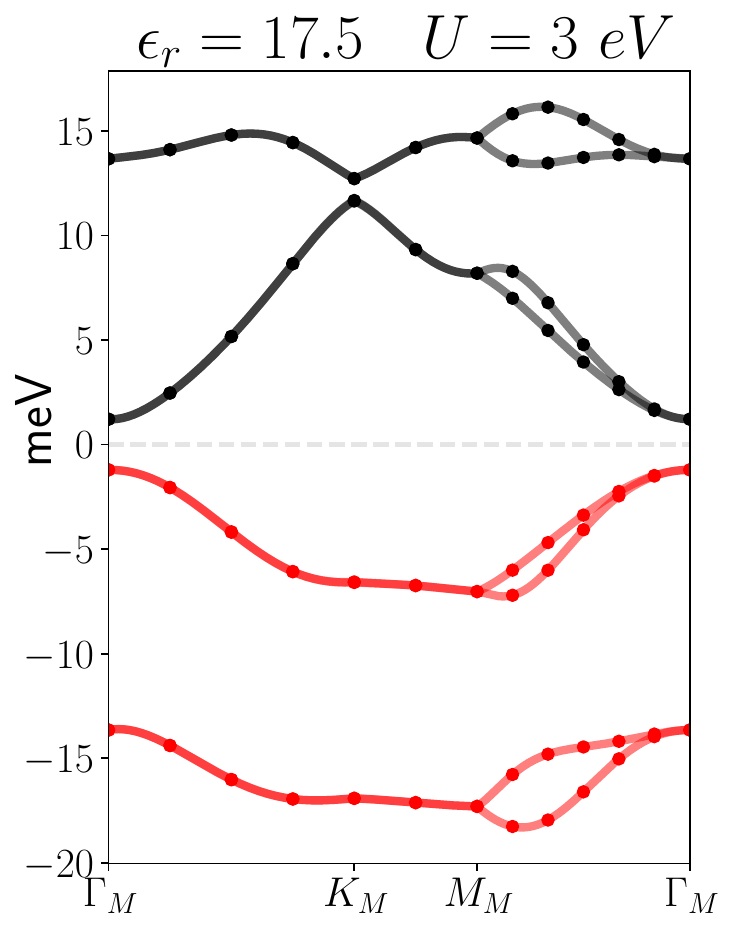}
    \includegraphics[width=.13\linewidth]{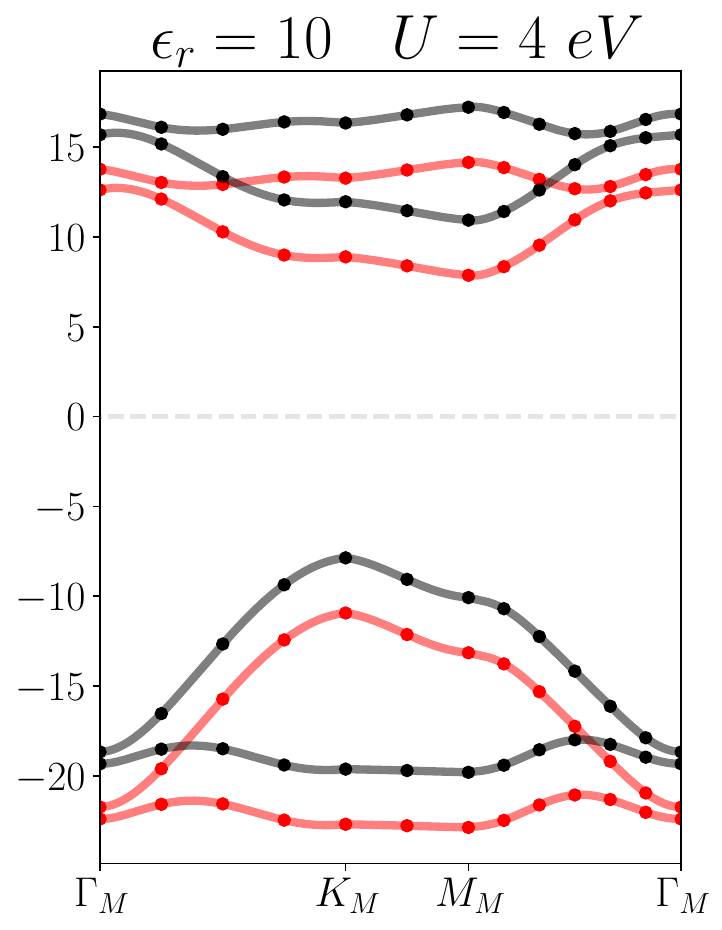}
    \includegraphics[width=.24\linewidth]{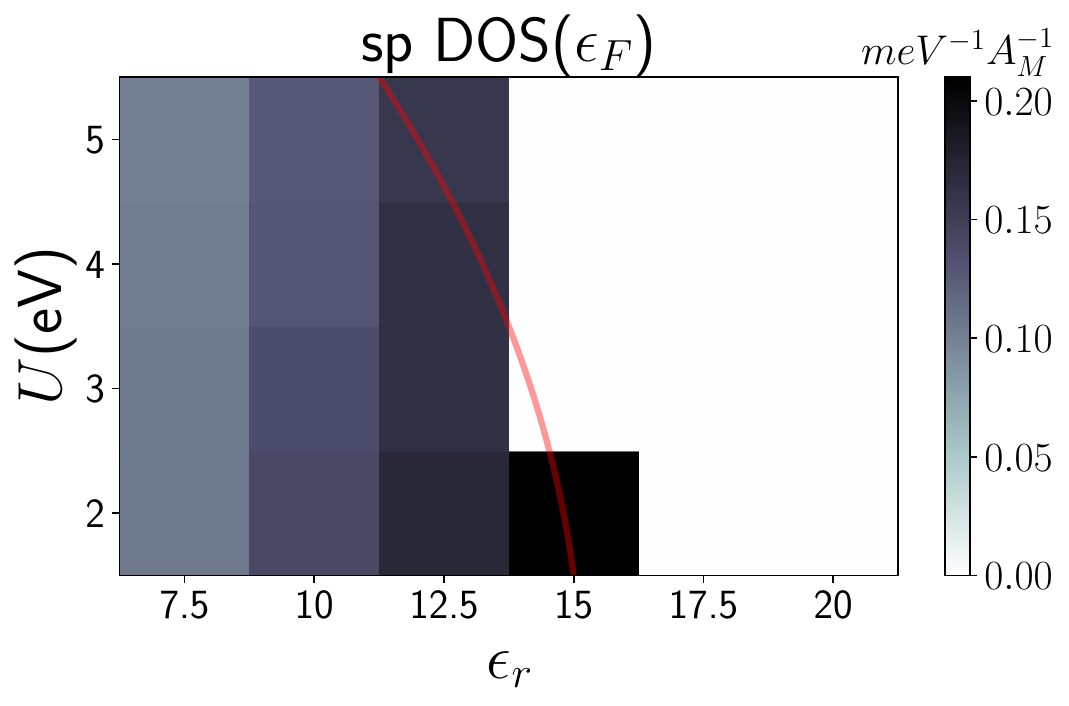}
    \includegraphics[width=.23\linewidth]{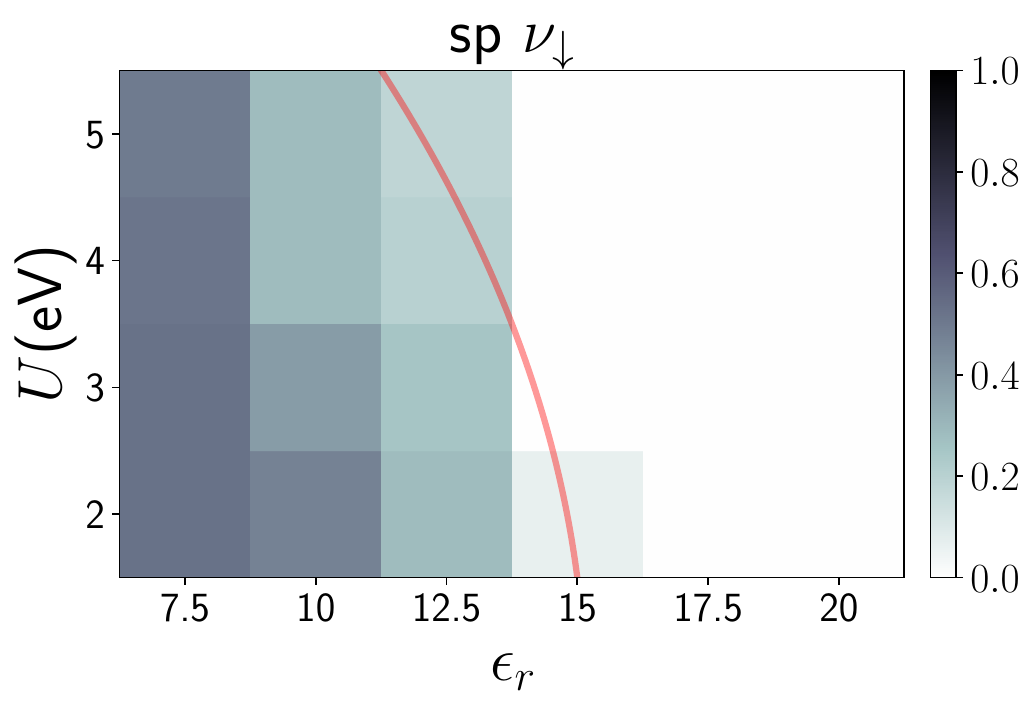}
    \includegraphics[width=.17\linewidth]{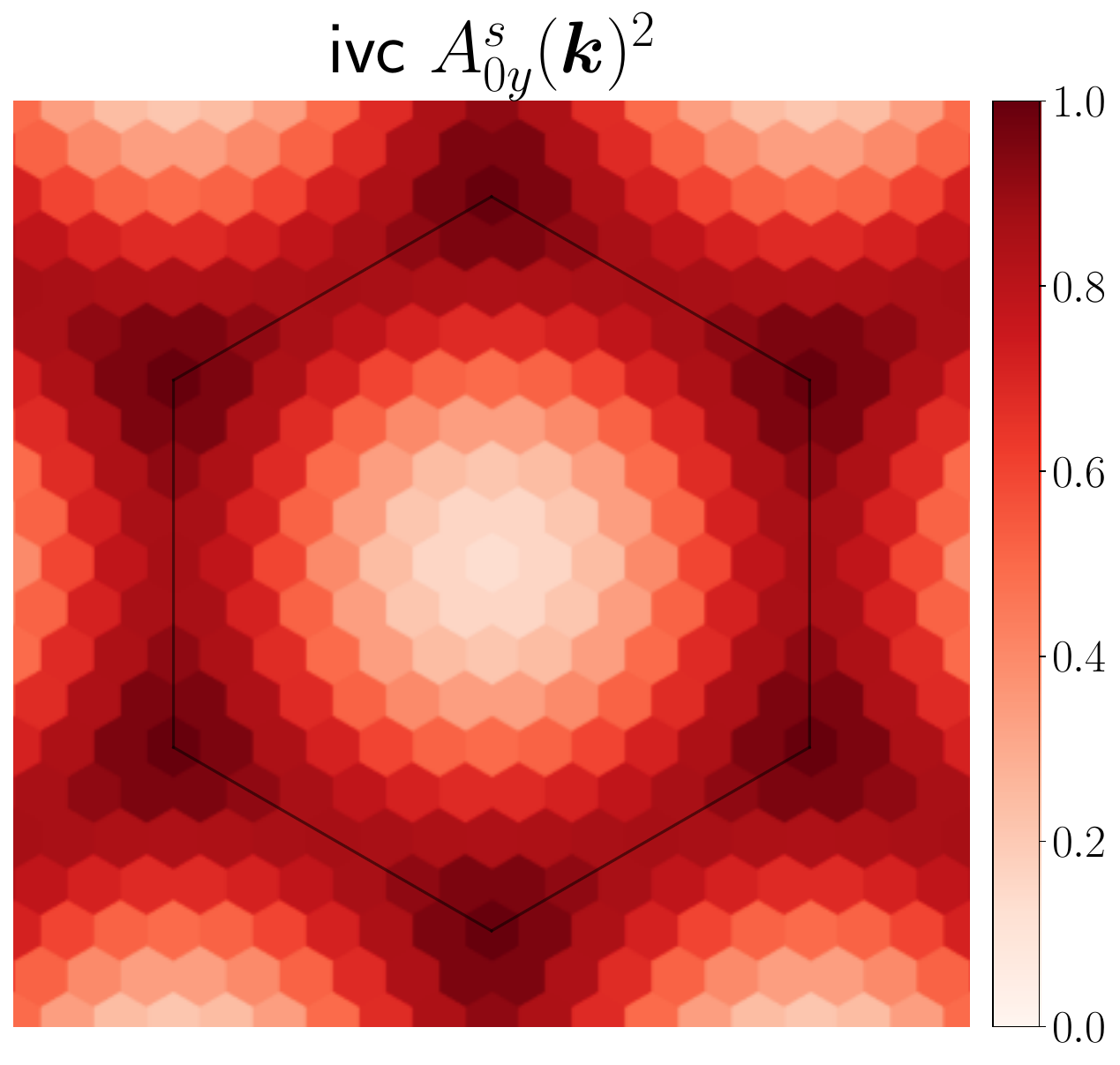}
\caption{\textbf{Competing states for $\boldsymbol{\nu=0}$.} \textbf{a)} Band structure of the spin polarized (sp) phase for $\epsilon_r=17.5$, $U=3$ eV (left) and the intervalley coherent phase for $\epsilon_r=10$, $U=4$ eV (right) \textbf{b)} Density of states at the Fermi level and spin down population of the sp phase as a function of $\epsilon_r$ and $U$. The red line is the tentative phase transition line, the sp phase being stable to the right. \textbf{c)} Intervalley coherent order parameter in the ivc phase for $\epsilon_r=10$, $U=4$ eV.}
\label{nu0hfplots}
\end{figure}

\begin{figure}[H]
  % \hspace{-0.2cm} $a)$ \hspace{5.3cm} $b)$ \hspace{5.3cm} $c)$ \hspace{4cm} \\
  \centering
  \includegraphics[width=.99\linewidth]{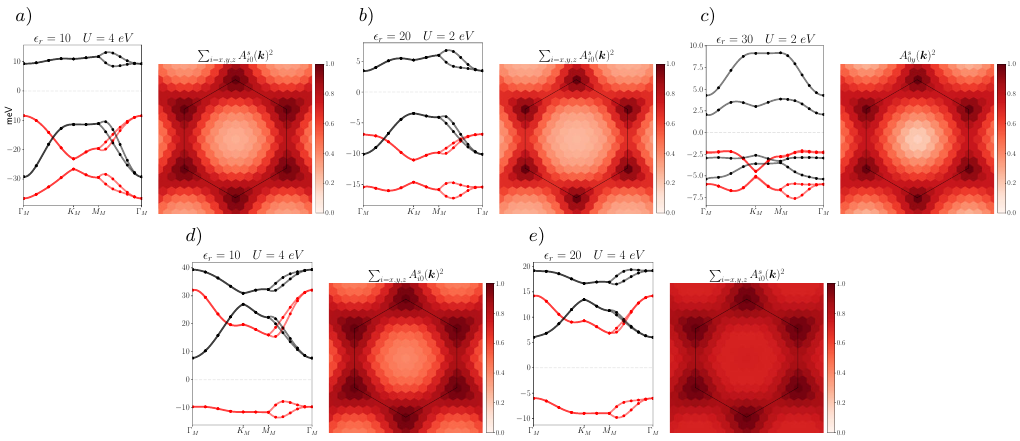}
  %  \includegraphics[width=.14\linewidth]{bande10u4n2.pdf}
  %  \includegraphics[width=.17\linewidth]{d0e10u4n2.pdf}
  %  \includegraphics[width=.14\linewidth]{bande20u2n2.pdf}
  % \includegraphics[width=.17\linewidth]{d0e20u2n2.pdf}
  % \includegraphics[width=.14\linewidth]{bande30u2n2.pdf}
  % \includegraphics[width=.17\linewidth]{divce30u2n2.pdf} \\
  %  \hspace{-5cm} $d)$ \hspace{5.3cm} $e)$ \hspace{5.3cm}\\
  % \includegraphics[width=.14\linewidth]{bande10u4n-2.pdf}
  % \includegraphics[width=.17\linewidth]{divce10u4n-2.pdf}
  % \includegraphics[width=.14\linewidth]{bande20u4n-2.pdf}
  % \includegraphics[width=.17\linewidth]{divce20u4n-2.pdf}\\
  \caption{\textbf{Band structures and main order parameter distributions} in the ground state for several selected interaction strengths. \textbf{a)} $\epsilon_r=10$, $U=4$ eV, $\nu=+2$. \textbf{b)} $\epsilon_r=20$, $U=2$ eV, $\nu=+2$. \textbf{c)} $\epsilon_r=30$, $U=2$ eV, $\nu=+2$. \textbf{d)} $\epsilon_r=10$, $U=4$ eV, $\nu=-2$. \textbf{e)} $\epsilon_r=20$, $U=4$ eV, $\nu=-2$.}
\label{nupm2hfplots}
\end{figure}

\begin{figure}[H]
  \centering
  \includegraphics[width=.6\linewidth]{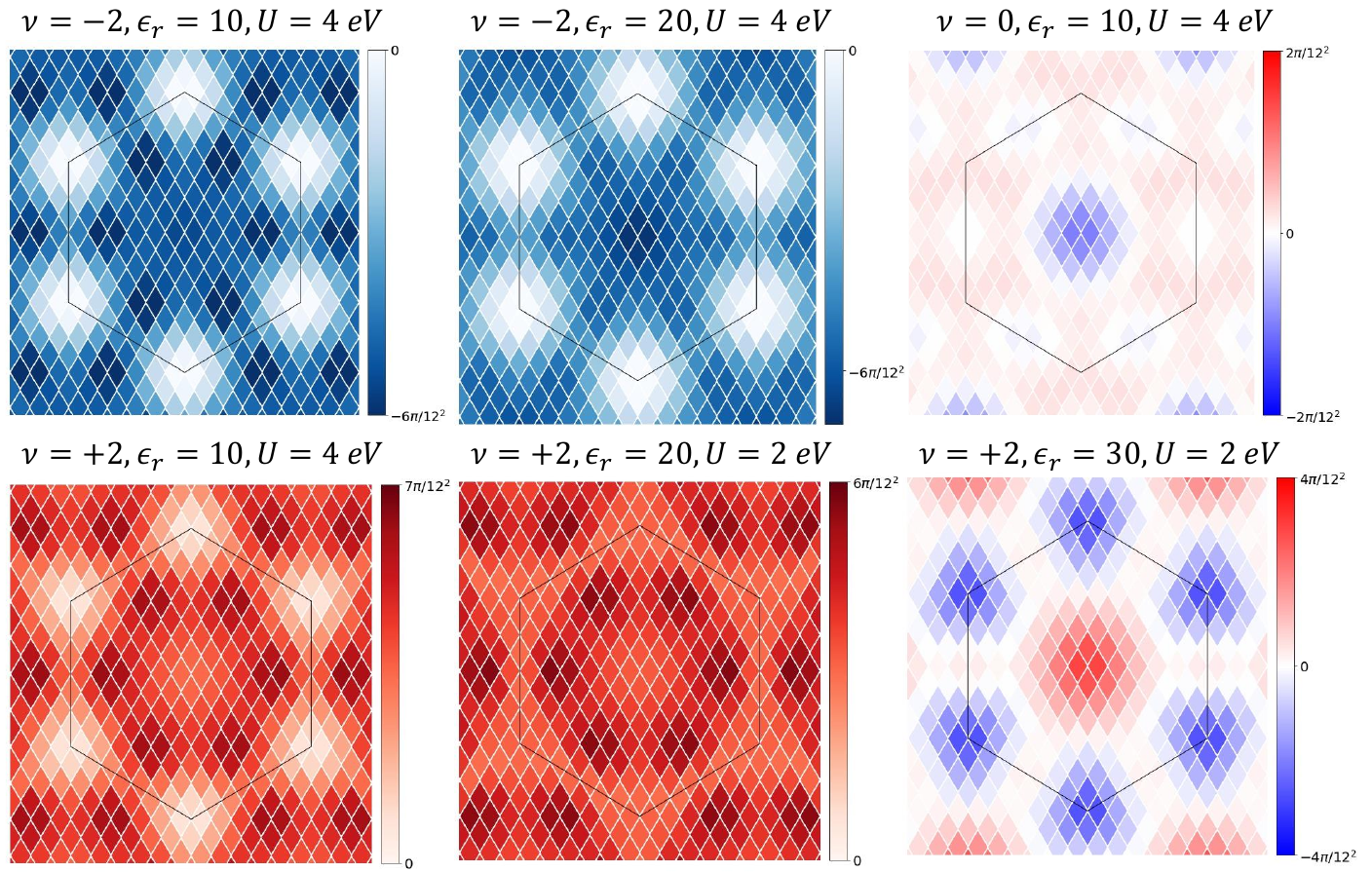}
  \caption{\textbf{Berry curvatures.} For several selected states, we obtain the non abelian Berry curvature from the projector onto the occupied flat bands. We plot the trace of the Berry curvature integrated on the parallelograms defined by the $12\times 12$ grid in the Brillouin zone. For $\nu=0$, we show the Berry curvature of one spin species (they are equal for both spins). The Chern numbers reported in the main text are reproduced after summing over the Brillouin zone.}
\label{berry}
\end{figure}

In Fig. \ref{nu0hfplots} we show additional results for both the intervalley coherent and spin polarized phases at $\nu=0$. In Fig. \ref{nupm2hfplots} we show band structures and order parameter distributions of several selected states, for $\nu=\pm 2$. Finally, in Fig. \ref{berry} we plot the Berry curvature distributions of several ground states.”

A salient feature of the band structures is the degeneracy along the $\Gamma_M K_M M_M$ line, that is then lifted along the $M_M \Gamma_M$ line. This can be explained by the crystallographic and $U(1)$ valley symmetries.

First, let us remind the reader that the symmetries $C_{2z}$ and $C_{2y}\mathcal{T}$ act as $C_{2z}(k_x,k_y) = (-k_x,-k_y)$ and $C_{2y}\mathcal{T} (k_x,k_y) = (k_x,-k_y)$. As such, $C_{2z}$ interchanges the valleys and $C_{2y}\mathcal{T}$ preserves the valleys, see Fig. \ref{bandsnonint}a). The symmetry $C_{3z}$ also preserves the valleys. 

Now, the line $\Gamma_M K_M$ is invariant under $C_{3z}C_{2z}C_{2y}\mathcal{T}$, see Fig. \ref{lattice}b), but this transformation changes the valley, so there will be two degenerate states with different valleys along this line. Also, the line $K_M M_M$ is invariant under $C_{2z}C_{2y}\mathcal{T}$ and the same argument applies. For the line $M_M \Gamma_M$ it is not possible to make such construction and the degeneracy is not enforced. 

Notice that the valley symmetry appears when we assign a valley charge to the eigenstates. If $U(1)_v$ is broken, like it is spontaneously in the ivc phase, the degeneracy is lifted, see Fig. \ref{nu0hfplots}a).

\end{document}